\documentclass[10pt,a4paper]{article}
\usepackage{amsmath,latexsym,amssymb,amsfonts}
\usepackage[dvips]{graphicx}
\usepackage{bm}

\begin{document}

\title{\textbf{Inside and outside stories of black-branes in\\ anti de Sitter space}}
\author{\textsc{Jakob Hansen}$^{a}$\footnote{hansen@kisti.re.kr},\;
\textsc{Bum-Hoon Lee}$^{b,c}$\footnote{bhl@sogang.ac.kr},\;
\textsc{Chanyong Park}$^{b,c}$\footnote{cyong21@gmail.com}\\
and \textsc{Dong-han Yeom}$^{c,d}$\footnote{innocent.yeom@gmail.com}\\
\textit{$^{a}$\small{KISTI, Daejeon 305-806, Republic of Korea}}\\
\textit{$^{b}$\small{Department of Physics, Sogang University, Seoul 121-742, Republic of Korea}}\\
\textit{$^{c}$\small{Center for Quantum Spacetime, Sogang University, Seoul 121-742, Republic of Korea}}\\
\textit{$^{d}$\small{Yukawa Institute for Theoretical Physics, Kyoto University, Kyoto 606-8502, Japan}}}
\maketitle

\begin{flushright}
{\tt YITP-13-59}
\end{flushright}

\begin{abstract}
In this paper, we investigate the dynamics inside and outside of black-branes in anti de Sitter space by numerical simulations using double-null formalism. We prepare a charged planar matter shell which, due to a negative cosmological constant, collapses and dynamically forms a black-brane with an apparent horizon, a singularity and a Cauchy horizon. The gravitational collapse cannot form a naked overcharged black-brane and hence weak cosmic censorship is safe. Although mass inflation occurs, the effect is much milder than in the case of charged black holes; hence, strong cosmic censorship seems not to be safe. We observed the scalar field dynamics outside the horizon. There should remain a non-trivial scalar field combination -- `charge cloud' -- between the horizon and the boundary. This can give some meaning in terms of the AdS/CFT correspondence.
\end{abstract}

\newpage

\tableofcontents

\newpage

\section{Introduction}

String theory predicts the existence of various solitonic objects, 
so-called D-branes \cite{Polchinski:1998rq}. If the D-branes contain enough energy, then 
they will form black objects called black-branes with a planar horizon and a singularity \cite{Callan:1996dv}.  
If one regards the string theory as a good candidate of the theory of everything, then the detailed study of black objects are very worthwhile. Moreover, there have been many recent works
paying attention to the black-brane geometry for understanding strongly interacting systems. 
In the context of AdS/CFT correspondence \cite{Maldacena:1997re,Gubser:1998bc,Witten:1998qj}, black-branes describe a bulk geometry dual to a quantum field theory of the dense medium. This will give fruitful results for application of conformal field theory, nuclear physics, condensed matter physics, and so on.

However, up to now, there have been limitations on the study the black objects, since many people relied on analytic techniques. Hence, what people usually have done has been to find static solutions study perturbations of it. Of course, the perturbation method is powerful but not exact and for higher accuracy
and deeper understanding of dynamical aspects of black-brane formation and evolution, 
a numerical approach is required. There are a lot of papers which study the dynamical formation 
of black holes and black-branes in order to understand its own gravitational features as well as
the time evolution of the non-thermal state, following the spirit of the AdS/CFT correspondence.

In this paper, we study dynamical formation, evolution and response of scalar perturbations 
of black-branes, by using full numerical simulations employing double-null formalism 
\cite{Hamade:1995ce,doublenull,Hansen:2009kn,Hong:2008mw,Hwang:2011mn,Hwang:2010aj,Hwang:2012pj}. 
To do so, we consider an Einstein-Maxwell-scalar model in which the scalar field is complex
and has an U(1) charge. A similar model, in which the mass square of a complex scalar 
given by $-2$, was taken into account studying the holographic superconductivity where only the
static black-brane solution was considered \cite{Hartnoll:2008vx}. Related to the stability of such a model,
it would be interesting to investigate the time evolution of solutions which
may give rise to some information for the stabilization of the black-brane corresponding to the thermalization of its dual theory.
From this point on, we concentrate on the massless complex scalar case in which 
the complex scalar field describes the scalar operator with an U(1) global charge and conformal dimension $3$ \cite{Erlich:2005qh}. On the other hand, the bulk gauge field is dual to the matter density operator \cite{Lee:2009bya}. The thermalization of this system, even in the strong
coupling regime, can be represented by the stabilization of the black-brane
because the black-brane leads to a well-defined thermal system \cite{Kachru:2008yh}. 

Due to the different topology of black-branes from black holes, 
the collision of planar/hyperbolic energy pulses in the asymptotically flat space time
does not form black objects \cite{Hwang:2012pj}. However, this is not the case in the 
asymptotic anti de Sitter background. Since there already exist well-known black-brane solutions,
the dynamical formation of such a black object should be allowed. 
Therefore, it is interesting to apply the double-null formalism to black-brane formation
in spacetimes with nontrivial asymptotic geometry.
In this paper, we especially focus on the inside and outside of black-branes. For the inside, we investigate the dynamics of singularities, Cauchy horizon and the cosmic censorship issue \cite{wald}. For the outside, we perturb the scalar field and observe the effects of it. Then the response can be seen by the boundary observer in anti de Sitter space and hence this technique will be useful for further applications of AdS/CFT.

This paper is organized as follows; In Section~\ref{sec:mod}, we describe technical details of the double-null formalism applied to black-brane dynamics. In Section~\ref{sec:in}, we investigate the causal structure and comment on weak and strong cosmic censorship. In Section~\ref{sec:out}, we observe the responses of the scalar fields and give physical interpretations. In Section~\ref{sec:dis}, we summarize our results and give future applications and finally in the Appendix, we present convergence tests of the numerical code.

\section{\label{sec:mod}Model for double-null formalism}

We consider Einstein gravity with a complex scalar field and a negative vacuum energy \cite{Hawking:1973uf}:
\begin{eqnarray}
S = \int dx^{4} \sqrt{-g} \left[ \frac{1}{16\pi} R + \mathcal{L} - V_{0} \right],
\end{eqnarray}
where
\begin{eqnarray} \label{Lagrangian}
\mathcal{L} = - \frac{1}{2}\left(\phi_{;a}+ieA_{a}\phi\right)g^{ab}\left(\overline{\phi}_{;b}-ieA_{b}\overline{\phi}\right) - m^2 \bar{\phi} \phi -\frac{1}{16\pi}F_{ab}F^{ab}.
\end{eqnarray}
Here, $R$ is the Ricci scalar, $\phi$ is a complex scalar field, $A_{\mu}$ is a gauge field, $F_{ab}=A_{b;a}-A_{a;b}$, $V_{0}<0$ gives a constant negative vacuum energy and $e$ is the gauge coupling. This theory for $m^2 = -2$ describes the holographic superconductivity of
a condensed matter system in the strong coupling 
regime \cite{Hartnoll:2008vx}. In this paper, we will take into account the $m=0$ case and investigate
the dynamical formation, evolution and stabilization of a black-brane which,
following the AdS/CFT correspondence, describes a thermalization process of the dual theory.

We can derive Einstein equations and scalar field equations:
\begin{eqnarray}\label{eq:Einstein}
G_{\mu\nu} = 8 \pi T^{\phi}_{\mu\nu},
\end{eqnarray}
where
\begin{eqnarray}
\label{eq:T_Phi}
T^{\phi}_{\mu\nu} &=& \frac{1}{2}\left(\phi_{;a}\overline{\phi}_{;b}+\overline{\phi}_{;a}\phi_{;b}\right)
\nonumber \\
&& {}+\frac{1}{2}\left(-\phi_{;a}ieA_{b}\overline{\phi}+\overline{\phi}_{;b}ieA_{a}\phi+\overline{\phi}_{;a}ieA_{b}\phi-\phi_{;b}ieA_{a}\overline{\phi}\right)
\nonumber \\
&& {}+\frac{1}{4\pi}F_{ac}{F_{b}}^{c}+e^{2}A_{a}A_{b}\phi\overline{\phi}+\mathcal{L}g_{ab} -V_{0} g_{\mu\nu}
\end{eqnarray}
and
\begin{eqnarray}
0 &=& \phi_{;ab}g^{ab}+ieA^{a}\left(2\phi_{;a}+ieA_{a}\phi\right)+ieA_{a;b}g^{ab}\phi,\\
0 &=& \frac{1}{2\pi}{{F^{b}}_{a}}_{;b}-ie\phi\left(\overline{\phi}_{;a}-ieA_{a}\overline{\phi}\right)+ie\overline{\phi}\left(\phi_{;a}+ieA_{a}\phi\right).
\end{eqnarray}

In this paper, we assume the double-null metric ansatz \cite{Hansen:2009kn,Hong:2008mw,Hwang:2011mn,Hwang:2010aj,Hwang:2012pj}:
\begin{eqnarray}\label{eq:ds}
ds^{2} = -2 e^{2\sigma(u,v)} du dv + r^{2}(u,v) d\Omega_{\kappa}^{2},
\end{eqnarray}
where we can impose symmetries as follows \cite{Hwang:2012pj}:
\begin{enumerate}
\item Spherical symmetry ($\kappa=+1$): $dS^{2} = d\theta^{2} + \sin^{2} \theta d\varphi^{2}$,
\item Planar symmetry ($\kappa=0$): $dR^{2} = dx^{2} + dy^{2}$,
\item Hyperbolic symmetry ($\kappa=-1$): $dH^{2} = d\chi^{2} + \sinh^{2} \chi d\varphi^{2}$.
\end{enumerate}

\subsection{Double-null formalism}

For simplification, we can choose a gauge fixing of the gauge field: $A_{\mu}= (a,0,0,0)$ \cite{doublenull,Hansen:2009kn,Hong:2008mw,Hwang:2011mn}. In addition, let us define
\begin{eqnarray}
\sqrt{4\pi}\phi \equiv s.
\end{eqnarray}
%and use conventions
%\begin{eqnarray}\label{eq:conventions}
%h \equiv \frac{\alpha_{,u}}{\alpha},\quad d \equiv \frac{\alpha_{,v}}{\alpha},\quad f \equiv r_{,u},\quad g \equiv r_{,v},\quad w \equiv s_{,u},\quad z \equiv s_{,v}.
%\end{eqnarray}

Then Einstein tensor components are
\begin{eqnarray}
\label{eq:Guu}G_{uu} &=& -\frac{2}{r} \left(r_{,uu}-2 r_{,u} \sigma_{,u} \right),\\
\label{eq:Guv}G_{uv} &=& \frac{1}{2r^{2}} \left( 4 r r_{,uv} + 2 \kappa e^{2\sigma} + 4r_{,u}r_{,v} \right),\\
\label{eq:Gvv}G_{vv} &=& -\frac{2}{r} \left(r_{,vv}-2 r_{,v} \sigma_{,v} \right),\\
\label{eq:Gthth}G_{aa} &=& -2 r^{2} e^{-2\sigma} \left(\sigma_{,uv}+\frac{r_{,uv}}{r}\right),
\end{eqnarray}
where $\kappa=+1, 0, -1$ and $a=\theta, x, \chi$ for spherical, planar, hyperbolic cases, respectively.
Energy-momentum tensor components are
\begin{eqnarray}
\label{eq:TPhiuu}T^{\phi}_{uu} &=& \frac{1}{4\pi} \left[ s_{,u}\overline{s}_{,u} + iea(\overline{s}_{,u}s-s_{,u}\overline{s}) +e^{2}a^{2}s\overline{s} \right],\\
\label{eq:TPhiuv}T^{\phi}_{uv} &=& \frac{{a_{,v}}^{2}}{8\pi}e^{-2\sigma} + e^{2\sigma} V_{0},\\
\label{eq:TPhivv}T^{\phi}_{vv} &=& \frac{1}{4\pi} s_{,v}\overline{s}_{,v},\\
\label{eq:TPhithth}T^{\phi}_{aa} &=& \frac{r^{2}}{8\pi}e^{-2\sigma} \left[ (s_{,u}\overline{s}_{,v}+s_{,v}\overline{s}_{,u}) + iea(\overline{s}_{,v}s-s_{,v}\overline{s})+{a_{,v}}^{2}e^{-2\sigma} \right] -r^{2}V_{0}.
\end{eqnarray}

Therefore, the simulation equations are as follows:
\begin{eqnarray}
\label{eq:E1}r_{,uu} &=& 2 r_{,u} \sigma_{,u} - 4 \pi r T^{\phi}_{uu},\\
\label{eq:E2}r_{,vv} &=& 2 r_{,v} \sigma_{,v} - 4 \pi r T^{\phi}_{vv},\\
\label{eq:E3}r_{,uv} &=& - \kappa \frac{e^{2\sigma}}{2r} - \frac{r_{,u}r_{,v}}{r} + 4\pi r T^{\phi}_{uv},\\
\label{eq:E4}\sigma_{,uv} &=& -\frac{4\pi e^{2\sigma}}{r^{2}}T^{\phi}_{aa} - \frac{r_{,uv}}{r}.
\end{eqnarray}
Additionally, we include the field equations:
\begin{eqnarray}\label{eq:S}
rs_{,uv}+r_{,u}s_{,v}+r_{,v}s_{,u}+iears_{,v}+iear_{,v}s+ies\frac{e^{2\sigma}q}{2r}&=&0,
\\
\left( \frac{r^{2}a_{,v}}{2}e^{-2\sigma} \right)_{,v}+\frac{ier^{2}}{4}(s_{,v}\overline{s}-s\overline{s}_{,v})&=&0,\\
\left( \frac{r^{2}a_{,v}}{2}e^{-2\sigma} \right)_{,u} - \frac{ier^{2}}{4}(s_{,u}\overline{s}-s\overline{s}_{,u}) + \frac{r^{2}}{2} e^{2} a s \bar{s} &=&0,
\end{eqnarray}
where we define the charge $q$ by
\begin{eqnarray}
q \equiv r^{2}a_{,v} e^{-2\sigma}.
\end{eqnarray}
In particular, the Maxwell equations can be rewritten as follows:
\begin{eqnarray}
\label{eq:m1}a_{,v} &=& \frac{e^{2\sigma}}{r^{2}}q,\\
\label{eq:m2}q_{,v} &=& -\frac{ier^{2}}{2}(s_{,v}\overline{s}-s\overline{s}_{,v}),\\
\label{eq:m3}q_{,u} &=& \frac{ier^{2}}{2}(s_{,u}\overline{s}-s\overline{s}_{,u}) - r^{2} e^{2} a s \bar{s},\\
\label{eq:m4}a_{,uv} &=& \frac{2e^{2\sigma}}{r^{2}} \left( \sigma_{,u}-\frac{r_{,u}}{r} \right) q + \frac{ie e^{2\sigma}}{2}(s_{,u}\overline{s}-s\overline{s}_{,u}) - e^{2\sigma} e^{2} a s \bar{s}.
\end{eqnarray}

\subsection{Initial conditions}

To give the proper initial conditions, we first define the concept of the mass function. We know that the general static solution for $V(\phi)=V_{0}$ will look like
\begin{eqnarray}
ds^{2} = - N(r)^{2} dt^{2} + \frac{{dr}^{2}}{N(r)^{2}} + r^{2} d\Omega_{\kappa}^{2},
\end{eqnarray}
with
\begin{eqnarray}
N^{2} = \kappa - \frac{2M}{r} + \frac{q^{2}}{r^{2}} - \frac{8\pi V_{0} r^{2}}{3}.
\end{eqnarray}
To connect this to the double-null coordinates, we use
\begin{eqnarray}        
dr &=& {r}_{,u}du + {r}_{,v}dv, \label{eq:a} \\
dt &=& \frac{e^{2\sigma}}{2} \left( -\frac{dv}{{r}_{,u}} + \frac{du}{{r}_{,v}} \right), \label{eq:b} 
\end{eqnarray}
and obtain the double-null metric $ds^{2} = -2 e^{2\sigma(u,v)} du dv + r^{2}(u,v) d\Omega_{\kappa}^{2}$.
Thus, we can show that
\begin{eqnarray}
N(r)^{2} = - 2{r}_{,u}{r}_{,v} e^{-2\sigma}.
\end{eqnarray}
Therefore, we can identify the local mass function as :
\begin{eqnarray}\label{eq:mass}
m(u,v) = \frac{r}{2} \left( 2 {r}_{,u} {r}_{,v} e^{-2\sigma} + \frac{q^{2}}{r^{2}} - \frac{8\pi V_{0}}{3} {r}^{2} \right).
\end{eqnarray}
Where $V_0$ is a negative constant (see subsequent sections). With this in mind, we need to specify initial conditions for all functions ($\sigma, r, s, a, q$), on the initial $u=0$ and $v=0$ surfaces.

At the starting point of our computational domain ($u=v=0$), we choose $\sigma_0=\sigma(0,0) = -\ln{ (1/\sqrt{2} )}$ so that the line element at this point becomes trivial (cf. Equation~(\ref{eq:ds})). Furthermore, at this initial point, we may choose $a(0,0)=q(0,0)=0$ as well as an initial mass of $m_0 = m(0,0)=0$. In principle we are free to choose an arbitrary intial gauge and hence we may chose a linear gauge with $r(0,0)=r_{0}=10$ and $r_{,u}(u,0) = \mathrm{const} < 0$ and $r_{,v}(0,v) = \mathrm{const} > 0$ on the initial null surfaces. However, to be consistent with Equation~(\ref{eq:mass}) and other variables, we cannot choose any arbitraty combination of $r_{,u}(0,0)$ and  $r_{,v}(0,0)$. We choose these values such that $|r_{,u}(u,0)| = |r_{,v}(0,v)|$ as ensure they are consistent with Equation~(\ref{eq:mass}). 

Finally, we may determine the other initial conditions:

\begin{description}
\item[In-going ($v=0$) null surface:]
On the ingoing null surface we choose $s(u,0)=0$. To determine $\sigma$ along this surface, we can now integrate Equation~(\ref{eq:E1}) to obtain $\sigma(u,0)$. We may now in principle integrate Equation~(\ref{eq:m3}) to obtain $q$ along the ingoing surface, however, since we choose the scalar field to be trivial along this surface, as a consequence, $q(u,0)$ becomes equally trivial (also so for $a(u,0)$). This provides us with values for all variables along the initial ingoing null surface.

\item[Out-going ($u=0$) null surface:]
On the outgoing null surface, we choose
\begin{eqnarray}\label{eq:s_pulse}
s(0,v)=A \sin^{2} \left( \pi \frac{v-v_{0}}{\Delta v} \right) \exp \left( 2\pi i \frac{v-v_{0}}{\Delta v} \right)
\end{eqnarray}
for $v_{0} \leq v < v_{0}+\Delta v$, and otherwise, $s(v,0)=0$, $A$ is some amplitude. We may now obtain $\sigma(0,v)$ by integrating Equation~(\ref{eq:E2}). Similarly, we may integrate Equations~(\ref{eq:m1}) and (\ref{eq:m2}), to determine values for $a$ and $q$ along the initial outgoing null surface. This finishes the assignments of the initial conditions.
\end{description}

These initial conditions correspond to a situation where the initial background geometry is a pure anti de Sitter space
with an initial charged scalar pulse, which propagates to the center.  
The nontrivial charge and energy density of the 
initial pulse can disturb the initial geometry and deform it to another solution. 
Here, by using numerical simulations, we will investigate the dynamical evolution of
the anti de Sitter space to a charged black-brane and the stabilization of it, which can be
reinterpreted as a thermalization of the dual matter theory according to the AdS/CFT correspondence.

\section{\label{sec:in}Inside story}

In this section, we study the inside of black-branes during dynamical gravitational collapses. We fix the initial pulse by $A=0.1$, $v_{0}=10$, $\Delta v = 10$. We vary the gauge coupling $e$ and the vacuum energy $V_{0}$ and compare the results.

\subsection{Causal structures}

\begin{figure}
\begin{center}
\includegraphics[scale=0.4]{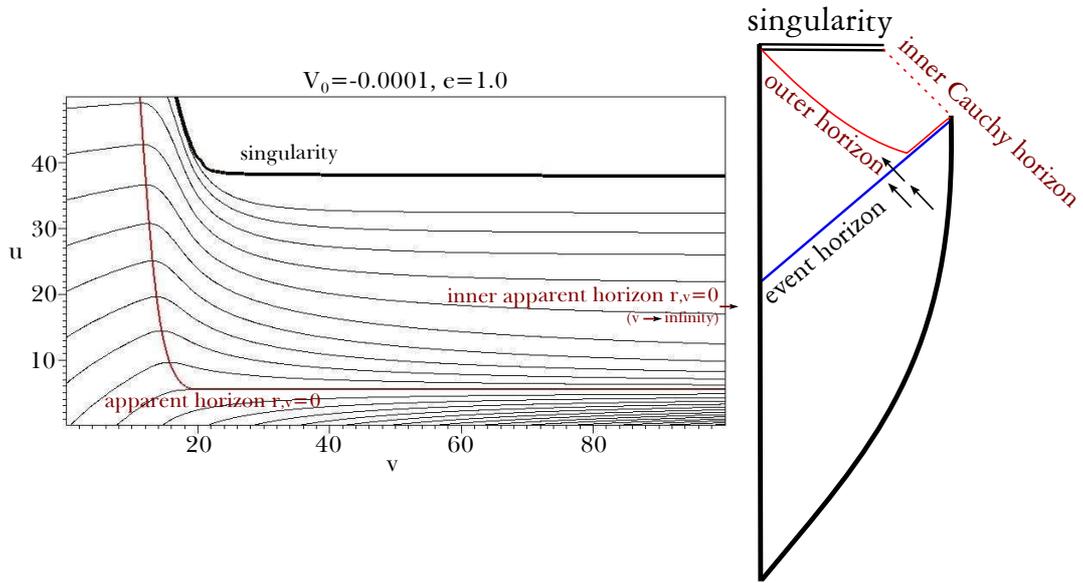}
\caption{\label{fig:causal}Causal structure of charged black-branes.}
\end{center}
\end{figure}

\begin{figure}
\begin{center}
\includegraphics[scale=1]{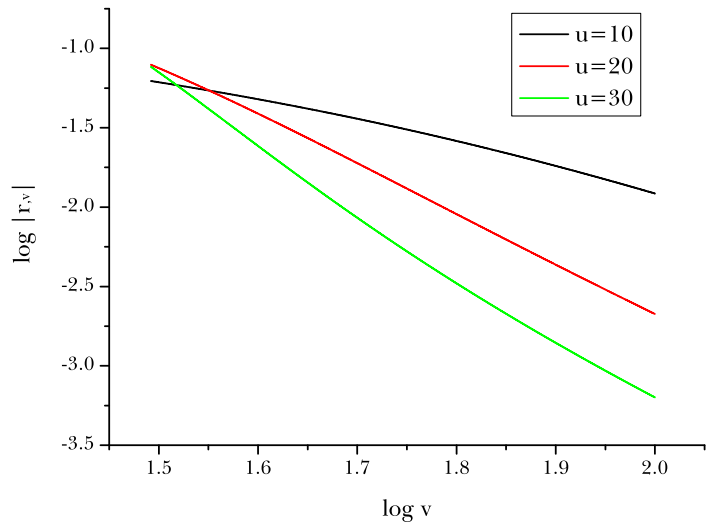}
\caption{\label{fig:G}$r_{,v}$ along $u=10, 20, 30$ slices. This can be approximated $r_{,v}\simeq v^{-B}$, where $B=1.43, 3.15, 4.07,$ respectively.}
\end{center}
\end{figure}

The left side of Figure~\ref{fig:causal} is the typical causal structure of a charged black-brane. Here, we used $V_{0}=-0.0001$ and $e=1.0$. After the gravitational collapse, the apparent horizon $r_{,v}=0$ grows in a space-like manner and quickly approaches an asymptote along the out-going null direction. Deep inside the black-brane, we can see a space-like $r=0$ singularity. Inside of the (apparent) horizon, as $v$ increases, one can see the fall-off of $r_{,v} \sim v^{-p}$ with $p> 1$ (Figure~\ref{fig:G}). This shows that in the $v\rightarrow \infty$ limit, both $r_{,v} \rightarrow 0$ and $r$ approaches a non-zero finite value. Therefore, this becomes an inner null horizon, that is the same as a spherical charged black hole. This causal structure can be summarized in the right diagram of Figure~\ref{fig:causal}. This is quite consistent with black hole cases \cite{Hong:2008mw,Hwang:2011mn}.

\begin{figure}
\begin{center}
\includegraphics[scale=0.29]{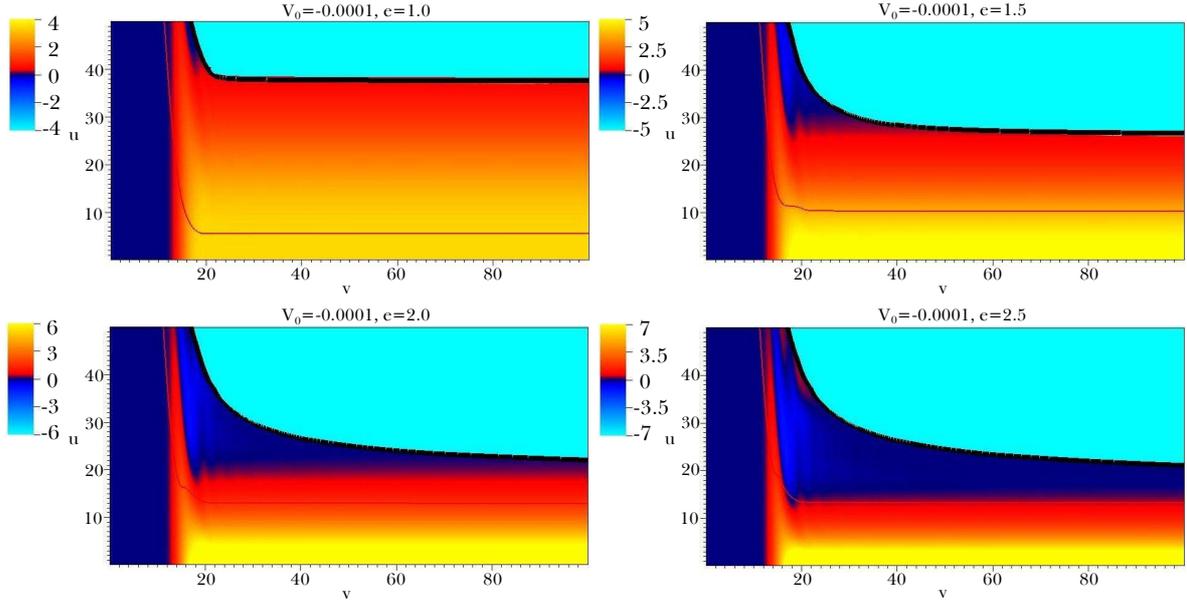}
\caption{\label{fig:L0001_Q}Causal structures for $V_{0}=-0.0001$ and varying of the gauge coupling $e=1.0, 1.5, 2.0, 2.5$. The color denotes the charge function $q$.}
\end{center}
\end{figure}

In Figure~\ref{fig:causal}, the black-brane is stabilized after the time $v=20$, at which the value of $r$ indicates the event horizon. Since $du=0$ and $r_{,v} = 0$ along the curve after this critical point, one can easily see from Equations~(\ref{eq:a}) and (\ref{eq:b}) that there is no variation in the radial coordinate. This fact implies that the increase of $v$ along the curve is nothing but the time evolution of the fixed event horizon. As a result, an initial charged scalar pulse deforms the initial anti de Sitter space to the black-brane geometry, which is stabilized after a finite time interval\footnote{We can estimate the time $N dt$ using Equation~(\ref{eq:b}) and it is approximately $\sim 51.1$.}. The resulting black-brane is charged due to the nontrivial distribution of the charge function (see Figure~\ref{fig:L0001_Q}).
If applying the AdS/CFT correspondence, the stabilization of the black-brane represents the thermalization of the $2+1$-dimensional dual field theory with matter, where the bulk gauge field is dual to the density operator 
while the charged scalar field in the asymptotic anti de Sitter space describes the condensate of the scalar operator with
a global U(1) charge and a conformal dimension of $3$.

By varying the gauge coupling $e$, one can see the robustness of the causal structures (Figure~\ref{fig:L0001_Q}). All diagrams share the same space-like apparent horizon, space-like $r=0$ singularity and inner null Cauchy horizon.

\begin{figure}
\begin{center}
\includegraphics[scale=0.4]{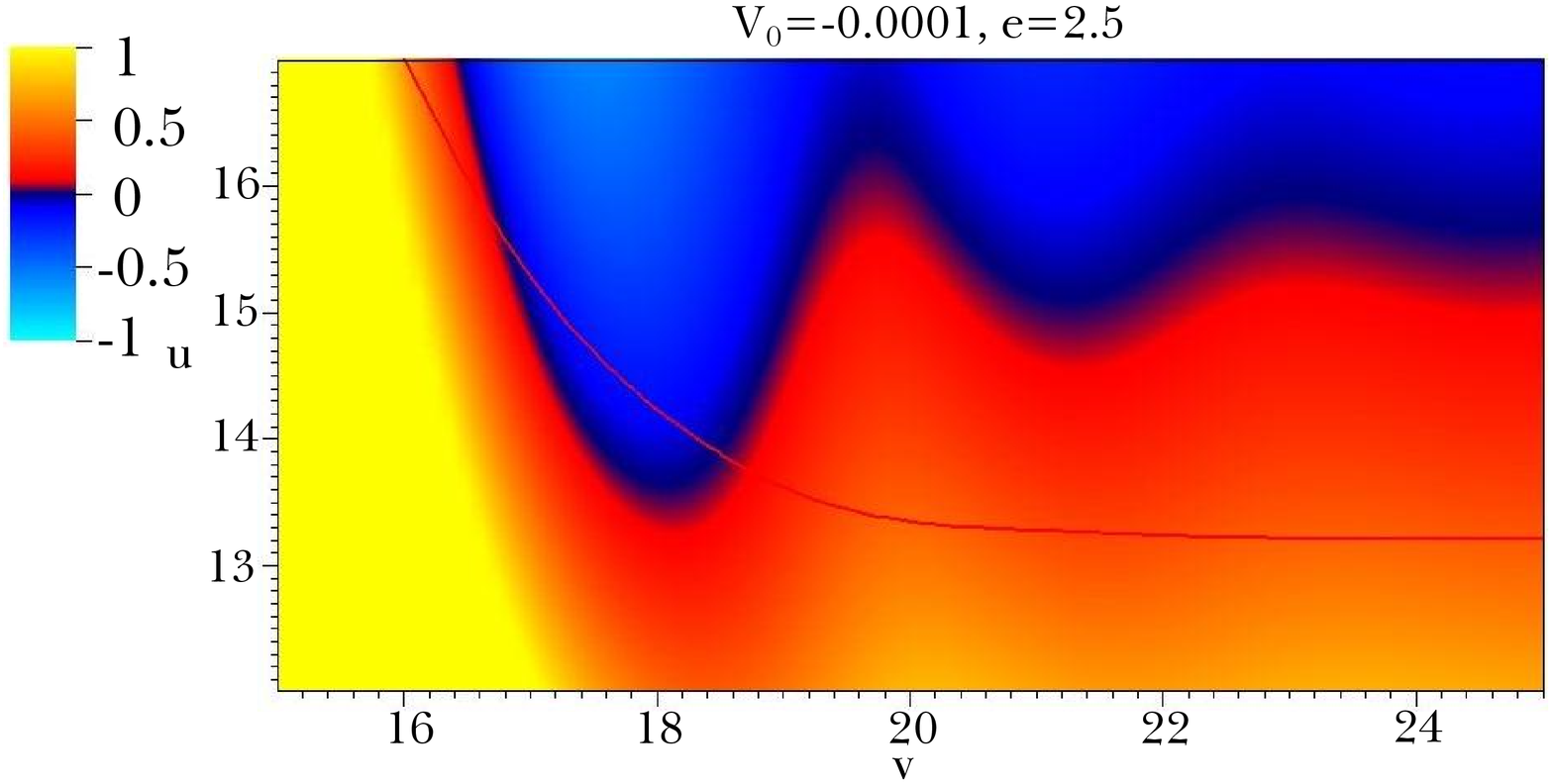}
\includegraphics[scale=0.7]{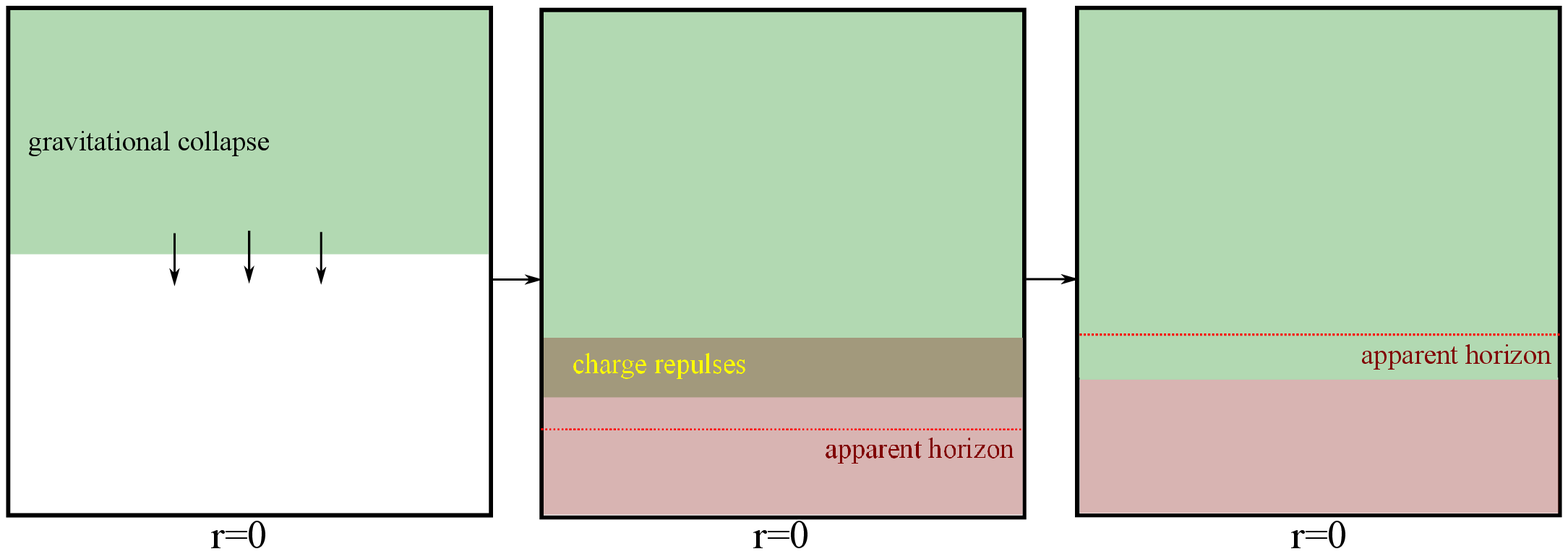}
\caption{\label{fig:discharging}Time dependent charge distribution of collapsing charged black branes. Upper: Region near horizon for $V_{0}=-0.0001$ and $e=2.5$. Lower: Schematic interpretation.}
\end{center}
\end{figure}

However, as $e$ increase, the asymptotic charge increase and the spatial distribution of charge changes. After sufficient time has passed (large $v$), the charge distribution approaches a stationary limit. However, before that happens, one can see interesting behavior. Upper part of Figure~\ref{fig:discharging} shows a region in the vicinity of the horizon for a large gauge coupling limit ($e=2.5$). The charge function at the horizon is initially positive but due to the field dynamics its sign changes, and eventually approaches a stationary value. If $e$ increases further, such charge oscillations become dominant. We can interpret this as follows (Lower part of Figure~\ref{fig:discharging}); First, the gravitational collapse happens (first figure). However, during the gravitational collapse, via large $e$ and large charge repulsion, locally some inside region becomes uncharged or even oppositely charged (second figure). Such an oppositely charged region can appear outside of the horizon. After sufficient time passed, the charge distribution will approach a stationary state (third figure), however, still some part inside of the horizon can be oppositely charged.

\subsection{Charge-mass relation and weak cosmic censorship}

Let us see whether the weak cosmic censorship is safe or not in the limit of large charge. The static black-brane solution becomes:
\begin{eqnarray}\label{eq:staticsol}
ds^{2} = - f(r) dt^{2} + f^{-1}(r) dr^{2} + r^{2} \left(dx^{2}+dy^{2}\right),
\end{eqnarray}
where
\begin{eqnarray}
f(r) = - \frac{2M}{r} + \frac{Q^{2}}{r^{2}} - \frac{8\pi V_{0}}{3}r^{2}.
\end{eqnarray}
This metric has two horizons (inner and outer). As the charge $Q$ increase, these horizons will approach each other and in the extreme limit, the two horizons will coincide. If the charge increase more than the extreme limit, then the solution will be a naked singularity and will violate the weak cosmic censorship.

\begin{figure}
\begin{center}
\includegraphics[scale=1]{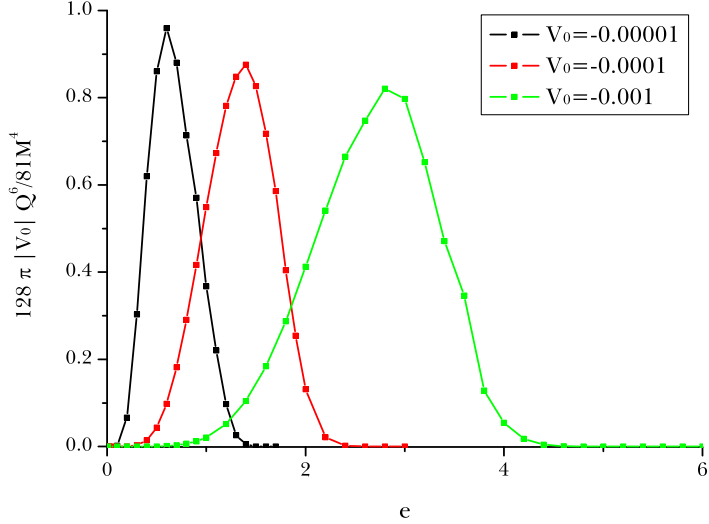}
\caption{\label{fig:q_m_e}$128\pi\left|V_{0}\right|Q^{6}/81M^{4}$ as function of by $e$ and $V_{0}$. Here, $Q$ and $M$ are observed on the horizon for sufficiently large $v$.}
\end{center}
\end{figure}

Let us find the condition for the extreme limit. The horizon $r_{0}$ for the extreme limit should satisfy two conditions: $f(r_{0}) = 0$ and $df/dr|_{r_{0}} = 0$. Or, equivalently,
\begin{eqnarray}
0 &=& - \frac{2M}{r_{0}} + \frac{Q^{2}}{r_{0}^{2}} - \frac{8\pi V_{0}}{3}r_{0}^{2},\\
0 &=& \frac{2M}{r^{2}_{0}} -2 \frac{Q^{2}}{r_{0}^{3}} - \frac{16\pi V_{0}}{3}r_{0}.
\end{eqnarray}
Then, the horizon $r_{0}$ in the extreme limit is
\begin{eqnarray}
r_{0} = \frac{2Q^{2}}{3M}
\end{eqnarray}
and at the same time, the condition
\begin{eqnarray}
\frac{128 \pi \left| V_{0} \right| Q^{6}}{81 M^{4}} = 1
\end{eqnarray}
should be satisfied. Therefore, to satisfy weak cosmic censorship, we require
\begin{eqnarray}
\mathcal{R} \equiv \frac{128 \pi \left| V_{0} \right| Q^{6}}{81 M^{4}} \leq 1.
\end{eqnarray}

We wish to check whether such kind of weak cosmic censorship is satisfied in a \textit{dynamical way} during gravitational collapse or not. In Figure~\ref{fig:q_m_e}, we vary $V_{0}$ and $e$ and checked the ratio $\mathcal{R}$ on the outer apparent horizon after the charge distribution approaches the stationary state. Interestingly, if $e$ is too small or too large, then the ratio is quite small and hence it looks neutral. In other words, if $e$ is too large, then the repulsion dominates and hence it repulses almost all charges; this causes the large coupling limit to be neutral. In addition, we observe that as $|V_{0}|$ increases, the standard deviation and peak of $\mathcal{R}$ increase. The most important thing is that $\mathcal{R}$ always stay below a value of one, which confirms our expectation that the process of gravitational collapse, dynamically conserves the weak cosmic censorship of black-branes.

\begin{figure}
\begin{center}
\includegraphics[scale=1]{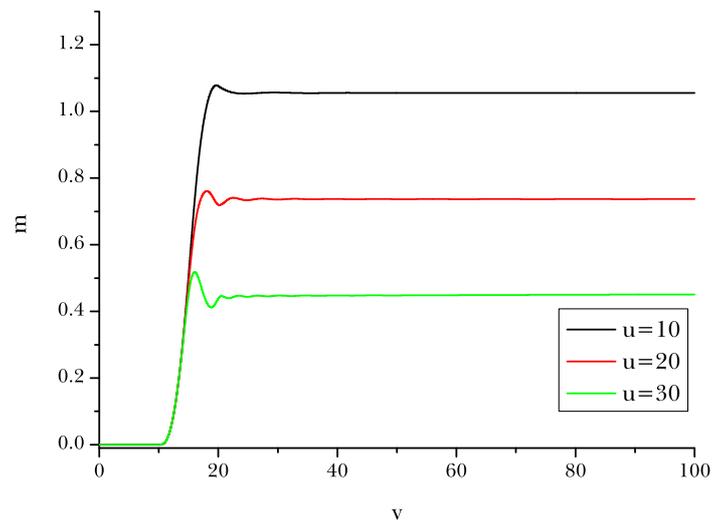}
\includegraphics[scale=1]{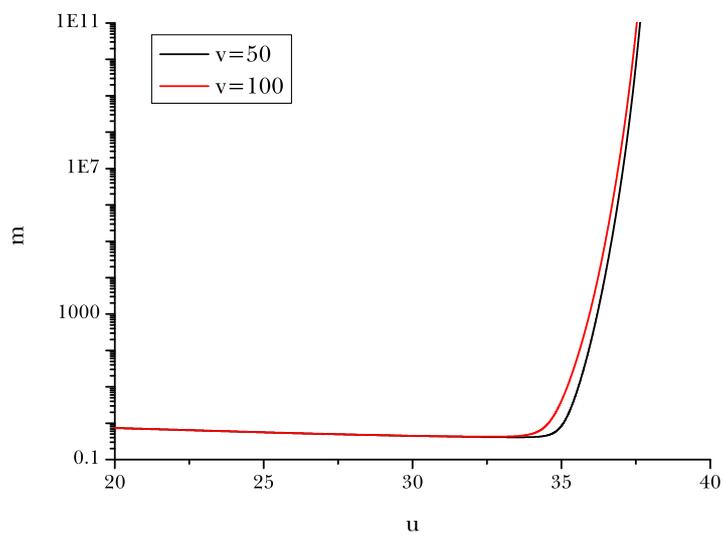}
\caption{\label{fig:mass}Mass function $m$ for $V_{0}=-0.0001$ and $e=1.0$ along some out-going and in-going null surfaces.}
\end{center}
\end{figure}

\subsection{Mass inflation and strong cosmic censorship}

Mass inflation is typical behavior near the Cauchy horizon in charged black objects \cite{Poisson:1989zz}.
We can re-write the metric in the following form:
\begin{eqnarray}
ds^{2} = - N^{2} dv^{2} + 2 dv dr + r^{2} d\Omega_{\kappa}^{2}.
\end{eqnarray}
Without loss of generality, one can ignore the symmetric part $\Omega_{\kappa}$ for the following calculations.

For in-falling matter along the in-going null direction (for coordinates $[v, r]$), the energy-momentum tensor components $T_{\alpha\beta}$ can be represented by
\begin{eqnarray}
T_{\alpha\beta} \propto \frac{F(v)}{r^{2}}\partial_{\alpha}v \partial_{\beta}v,
\end{eqnarray}
where $F(v)$ is an arbitrary function that refers to the luminosity function \cite{Poisson:1997my}. If there is an observer who moves along the out-going null direction and approaches the Cauchy horizon $r_{-}$, then for a null geodesic of the observer $l^{\alpha}$, $(\partial_{\alpha}v) l^{\alpha} = l^{v} = dv/d\eta$ and $dv/d\eta \propto \exp{\kappa_{-}v}$ near the $v\rightarrow \infty$ limit, where $\eta$ is an affine parameter of the observer and $\kappa_{-}$ is the surface gravity of the inner horizon. In conclusion, the out-going observer who approaches the Cauchy horizon measures the energy density for the energy flow that flows along the Cauchy horizon by \cite{Poisson:1997my}:
\begin{eqnarray}
\rho = T_{\alpha\beta}l^{\alpha}l^{\beta} \propto \frac{F(v)}{r^{2}} e^{2 \kappa_{-} v}.
\end{eqnarray}
For four dimensions and spherical symmetry, for a realistic black hole the behavior of the luminosity function should be polynomial to $F(v) \sim v^{-p}$ \cite{Price}. Therefore, it seems that any small energy flow along the Cauchy horizon is amplified for an observer who approaches to the Cauchy horizon. This effect is known as \textit{mass inflation}.

The previous description should be correct for general topologies, but the physical meaning is not entirely clear. For example, we do not have a general description of the luminosity function for the planar symmetry and hyperbolic symmetry in the anti de Sitter background. Also, if the topology is not spherical symmetry, then $r$ no longer represents the areal radius. However, it is fair to say that the assumptions of mass inflation for black hole cases are still reasonable. Therefore, it seems that mass inflation should be observed even for black-brane cases. If it is true, then the inner horizon cannot be stable and there should be strong back-reactions due to high energy density \cite{Ori}. Therefore, it is necessary to use numerical techniques to study these structures.

If one observer can penetrate beyond the null Cauchy horizon (of Figure~\ref{fig:causal}), then the observer may violate the strong cosmic censorship, unless the inner horizon becomes a curvature singularity. We observed the mass function along some out-going and in-going null surfaces (Figure~\ref{fig:mass}).

\begin{figure}
\begin{center}
\includegraphics[scale=0.5]{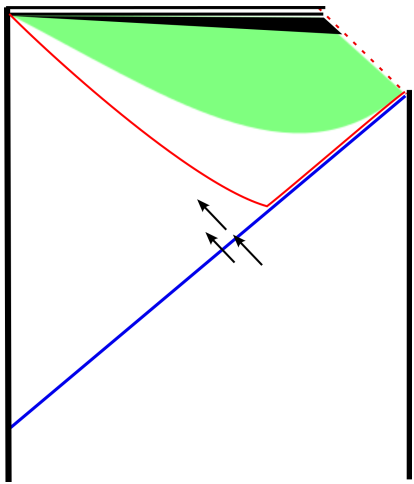}
\caption{\label{fig:minf}Schematic diagram of mass inflation for anti de Sitter black-branes and black holes. For the black hole case, mass inflation is observed around the green-colored region and hence the Cauchy horizon is covered \cite{Hansen:2009kn,Hong:2008mw,Hwang:2011mn}. For the black-brane case, the only black-colored region is protected by mass inflation.}
\end{center}
\end{figure}

The result is interesting; Along the out-going null direction, the mass function is quite stable. However, along the in-going null direction, the mass function becomes unstable once the coordinate $u$ increases beyond a critical limit. The critical $u$ corresponds to $r \simeq Q^{2}/2M$, where this is the approximate inner horizon radius of the static solution, Equation~(\ref{eq:staticsol}). Unless one observer approaches such a critical radius, the mass function is sufficiently small and no one can see mass inflation.

Such a behavior (when $r < Q^{2}/2M$, the mass begins to inflate) is consistent with the usual mass inflation \cite{Poisson:1989zz}. We can conclude that charged black-branes also have mass inflation as in black holes. However, there are quite distinct difference from the mass inflation of black holes (Figure~\ref{fig:minf}). For the black hole case, the appearance of exponential mass increase appears relatively sooner than for the black-brane cases. For black holes, once an observer falls beyond the event horizon, mass inflation is practically unavoidable for out-going observers. However, for black-brane cases, an observer inside of the event horizon can move along the out-going null direction and the observer can be safe without experiencing mass inflation. Therefore, for black-branes, it is highly non-trivial whether strong cosmic censorship will be satisfied or mass inflation will create a curvature singularity along the inner Cauchy horizon.

\section{\label{sec:out}Outside story}

\subsection{Scalar hair dynamics around the horizon}

\begin{figure}
\begin{center}
\includegraphics[scale=0.3]{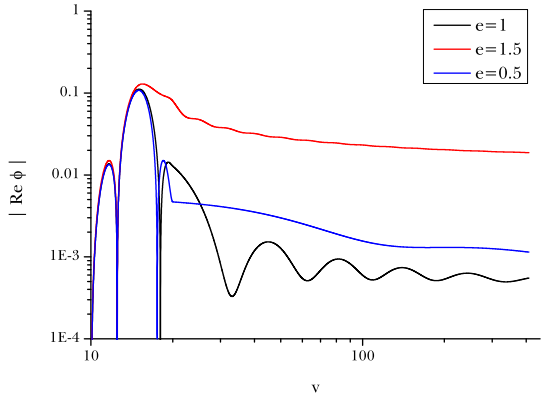}
\caption{\label{fig:hair2}The scalar field outside the black-brane without perturbation for $V_{0}=-0.0001$.}
\end{center}
\end{figure}

We first see the scalar field dynamics around the apparent horizon. Figure~\ref{fig:hair2} shows that as the gauge coupling increases, the scalar field decays more slowly and it seems to converge to a certain \textit{non-trivial value}. Perhaps, this can present a dynamical formation of a scalar hair around the black-brane.

Second, to perturb this scalar field, we slightly modify the initial scalar field after $v>v_{0}$ as follows:
\begin{eqnarray}
s'(0,v) = s(0,v) + B \frac{e^{\beta}}{(C_1-C_0)^{-\beta}}\frac{(v-C_0)^{-\beta}}{e^{\beta\frac{C_1-C_0}{v-C_0}}}.
\end{eqnarray}
This expression is designed to add a power-law tail to the primary scalar field pulse $s(0,v)$ (see Equation~(\ref{eq:s_pulse})), after the main pulse. The constant $C_0$ marks the beginning of the added tail (for $v<C_0$, we set $B=0$) and $C_1$ marks the beginning of the power-law decay, we usually set $C_0$ to coincide with the beginning of the main pulse and set $C_1$ to be inside the main pulse. This ensures that after the main pulse, we obtain a nice tail which decays as a power-law of order $\beta$. The expression is scaled so that the amplitude $B$ corresponds to the maximum of the added tail. We regard this as a scalar field perturbation from the boundary of the anti de Sitter to the black-brane. We hope to see the response of the scalar perturbation after an interaction with the black-brane. As long as $B$ is sufficiently small, it will not modify the general causal structure. We compared the scalar field dynamics with the $B=0$ case along the apparent horizon.

\begin{figure}
\begin{center}
\includegraphics[scale=0.3]{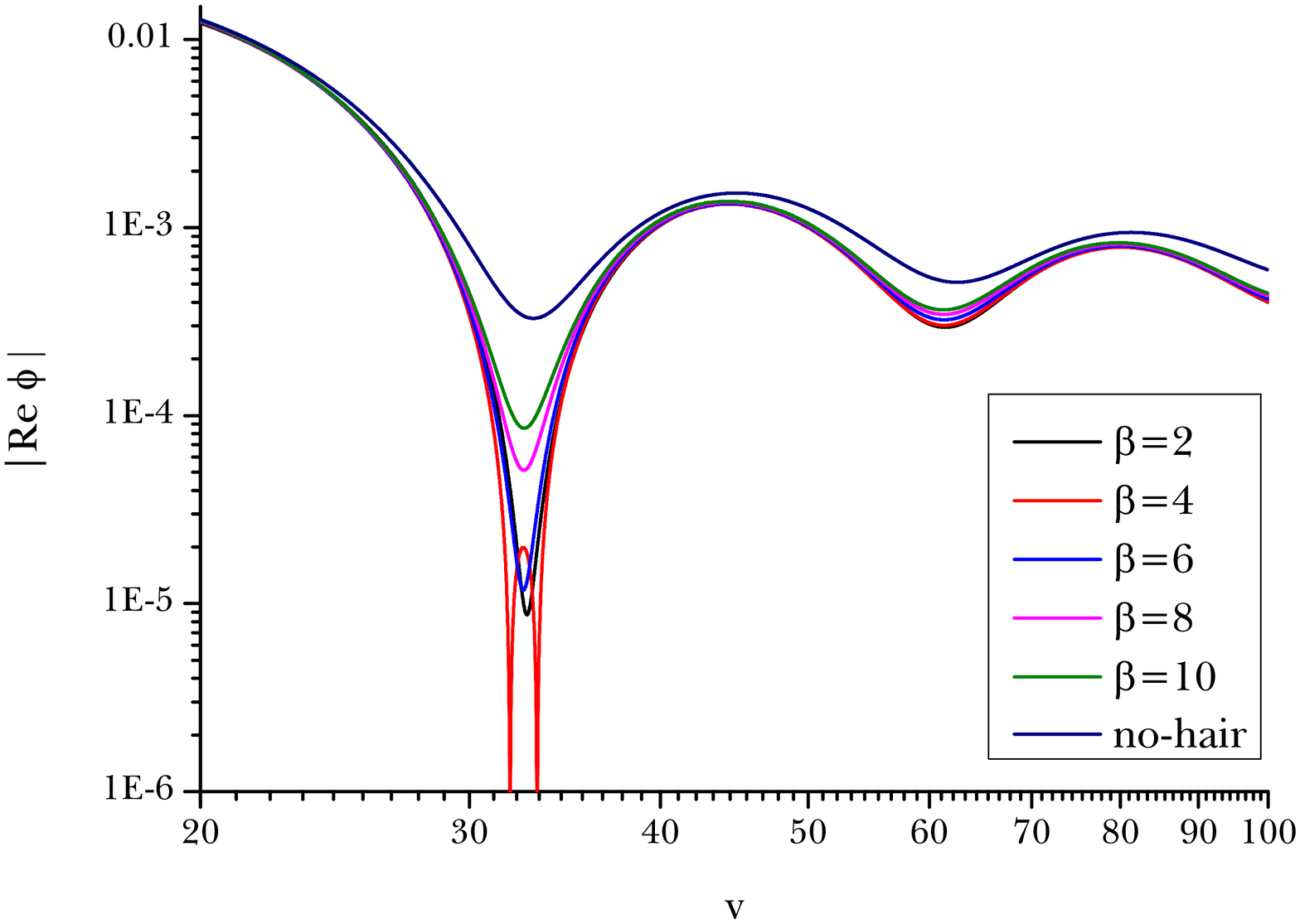}
\includegraphics[scale=0.3]{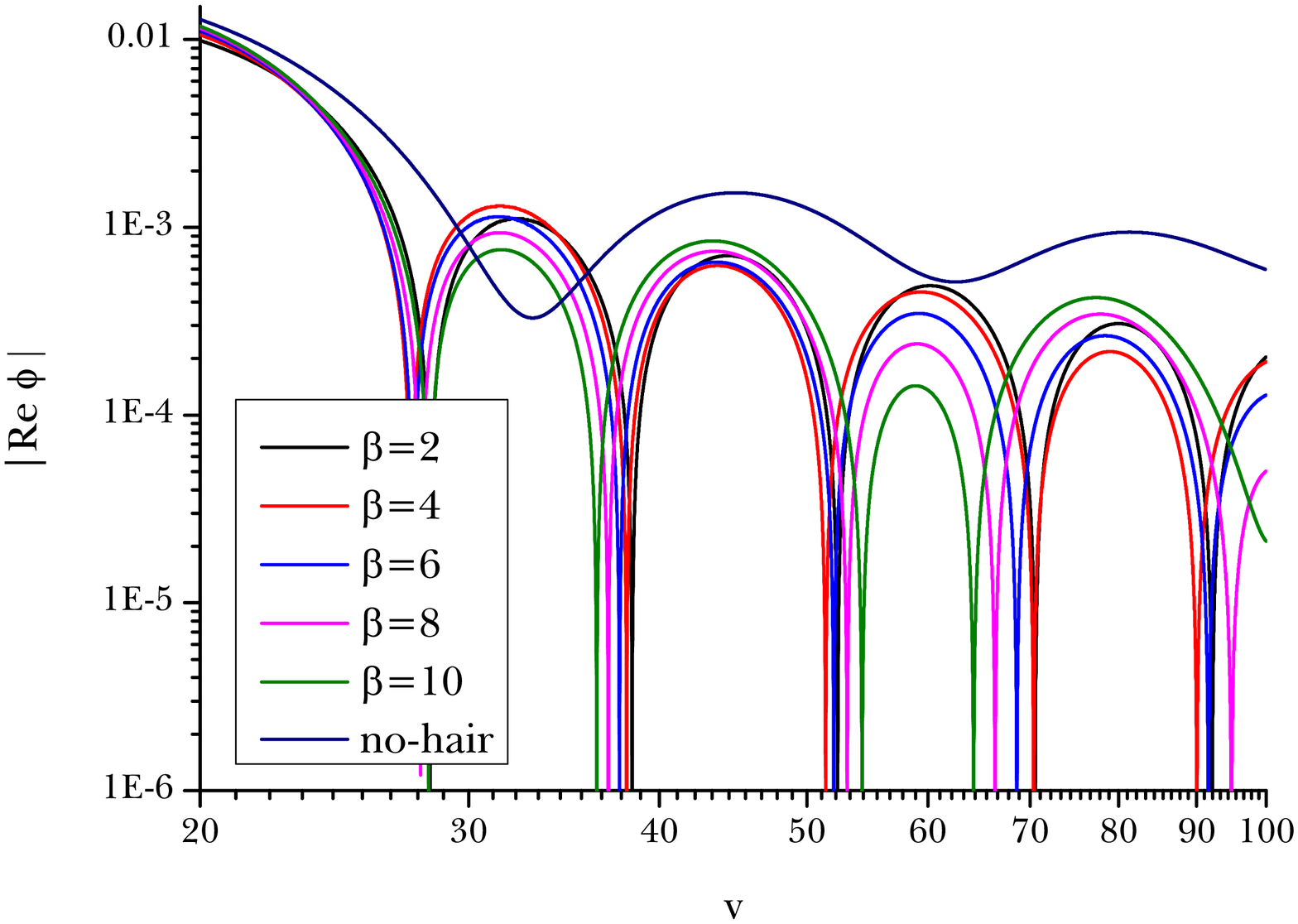}
\includegraphics[scale=0.3]{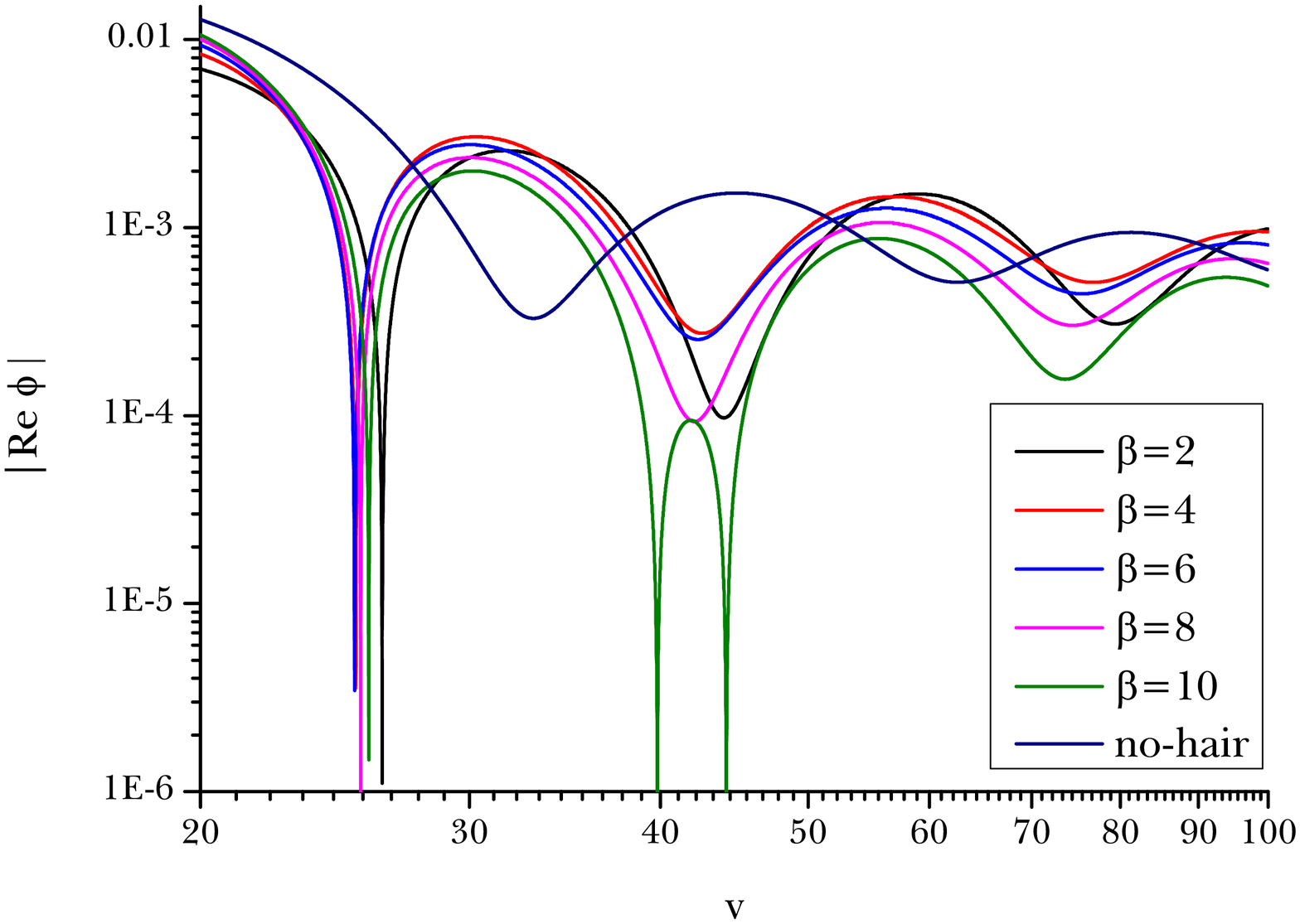}
\caption{\label{fig:B00001}Scalar field dynamics along the horizon for $V_{0}=-0.0001$ and $e=1.0$, by varying $B$ and $\beta$. Upper: $B=0.001$, Middle: $B=0.005$, and Lower: $B=0.01$.}
\end{center}
\end{figure}

Figure~\ref{fig:B00001} shows the scalar field dynamics along the apparent horizon. The typical behavior is that after a black-brane is formed, a part of the previous scalar field remnant is again absorbed into the black-brane, hence the black-brane slightly grows and then the horizon stabilizes after some oscillation. Here, the oscillation for the scalar field amplitude can be understood as the competition of the attractive gravitational and repulsive electric forces. It shows the damped oscillation behavior due to the friction caused by the curved background geometry. The rest of the scalar field remnant still remains outside the horizon due to the strong repulsive force. Such a repulsive force starts to prevent the next pulse from propagating to the center and causes the remnant of the scalar fields outside of the horizon.

\begin{figure}
\begin{center}
\includegraphics[scale=0.29]{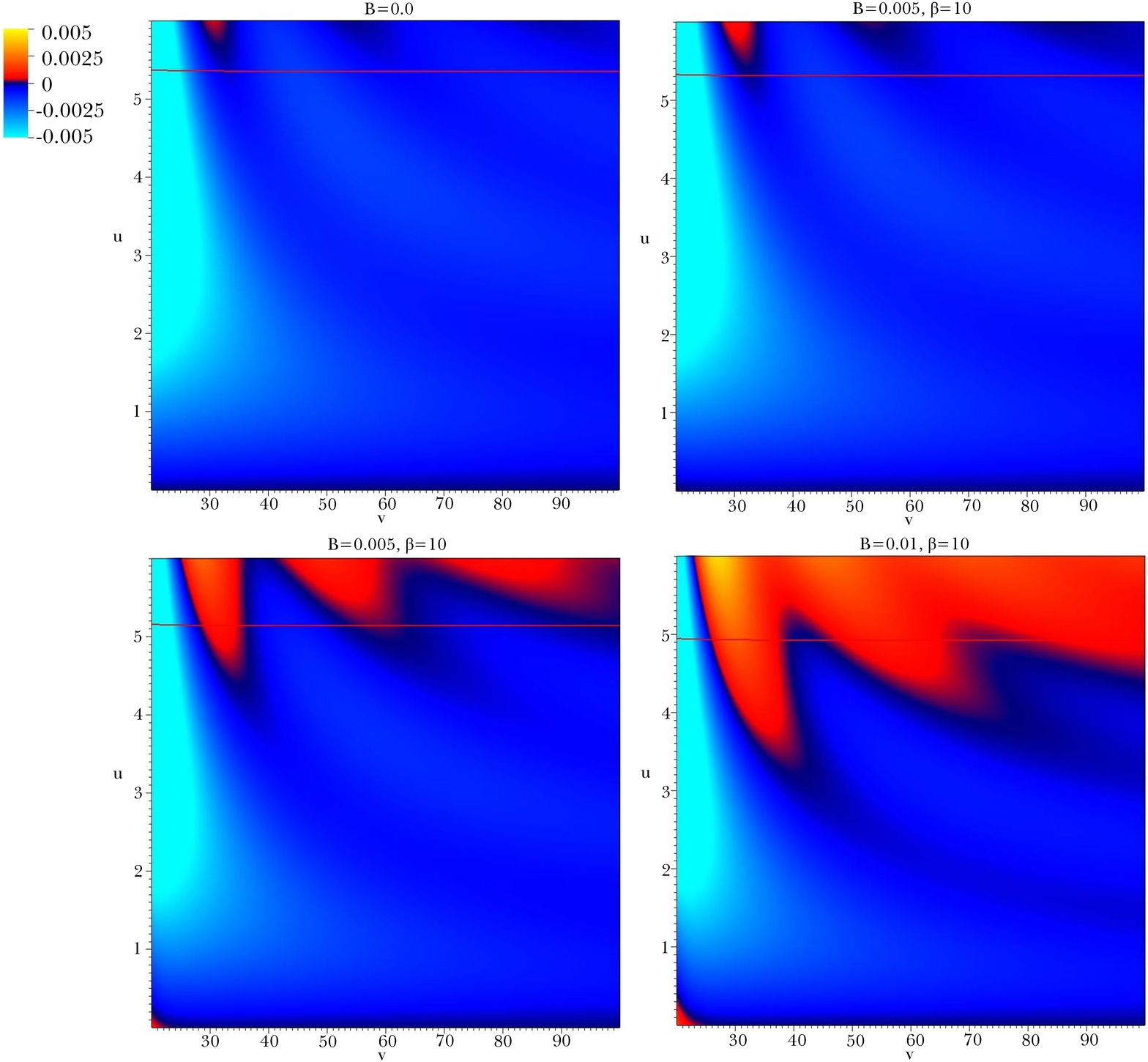}
\caption{\label{fig:hair}Plot of $\mathrm{Re} \;(s)$ with $V_{0}=-0.0001$ and $e=1.0$, by varying $B$ and $\beta$.}
\end{center}
\end{figure}

Figure~\ref{fig:hair} shows the overall picture of the scalar field. As the field amplitude increases, there are observable effects. The overall amplitude of the perturbed scalar field increases and the oscillating period decreases.

\subsection{Interpretations}

\begin{figure}
\begin{center}
\includegraphics[scale=0.4]{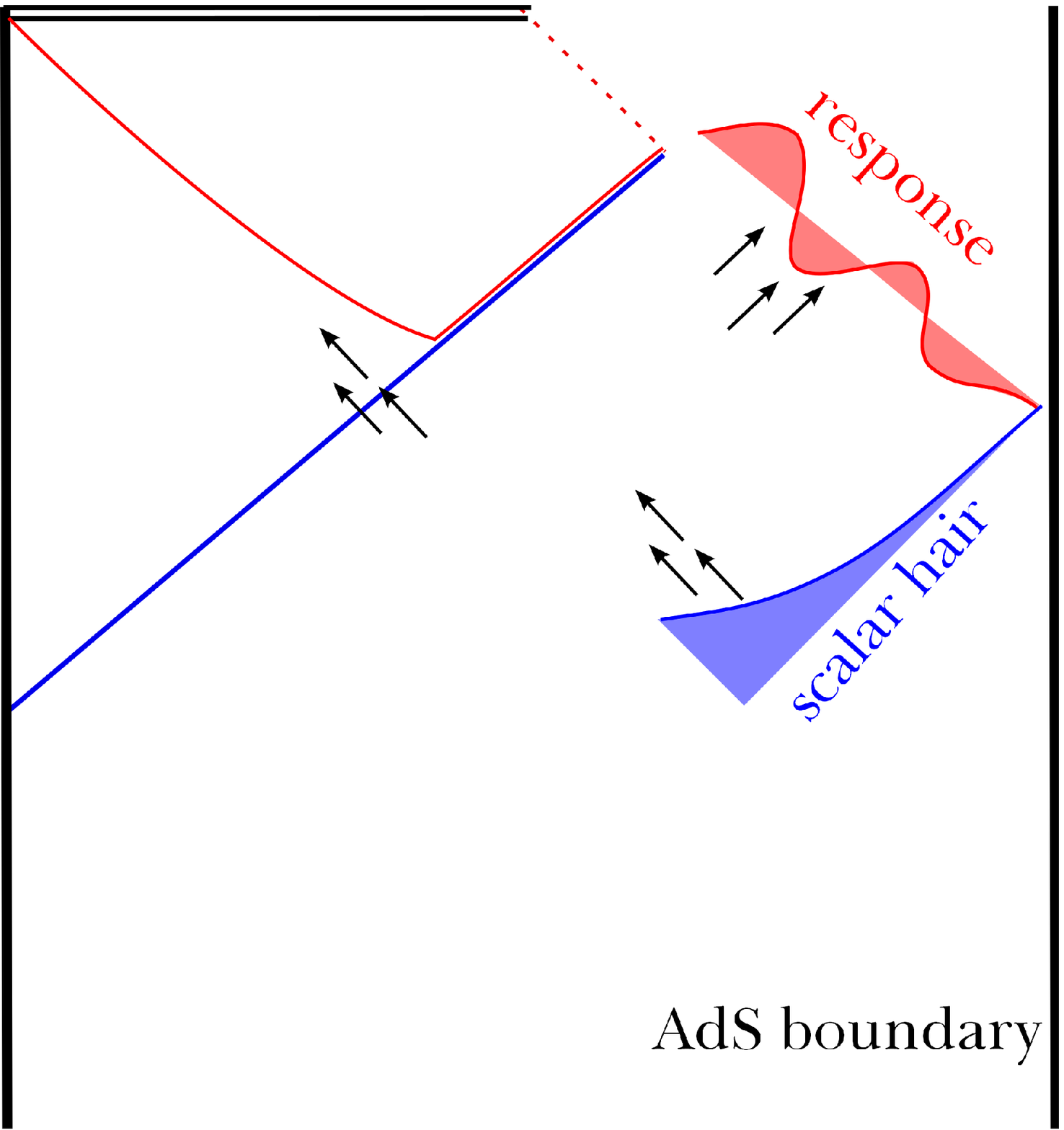}
\caption{\label{fig:interpretation}The response of the scalar field. The response should remain between the event horizon and the boundary.}
\end{center}
\end{figure}

One interesting result is that there \textit{remains} a non-trivial scalar field outside the event horizon (Figure~\ref{fig:hair2}). As in Figure~\ref{fig:interpretation}, this means that there is a non-trivial response to the out-going direction. What will happen for this response? Will it eventually collapse to the black-brane and disappear, or remain outside and approaches a stationary combination? It is not entirely clear for our simulations, since we could not observe the scattering of the scalar field from the boundary. However, we can interpret this way. Initially, we gave an \textit{over-charged} initial condition. Due to weak cosmic censorship, some of charge should be scattered. Then there will be an out-going scattered charge configuration. If the background is asymptotically de Sitter or Minkowski, then the scattered responses will disappear eventually and there will be no non-trivial scalar fields. On the other hand, if the background is anti de Sitter, then the response cannot escape to infinity. Hence, there should remain a \textit{charge cloud} between the event horizon and the boundary. This non-trivial scalar field configuration will remain forever, although we could not determine whether it approaches a stationary state or oscillates between the horizon and the boundary forever.

\section{\label{sec:dis}Discussion}

In this paper, we investigated the dynamics of inside and outside of a black-brane in anti de Sitter space, by numerical simulations using double-null formalism. In terms of numerical studies, it was interesting that a charged matter shell can form a black object without collision. This is due to the effect of anti de Sitter space.

We focused our attention to the inside and outside of black-branes. For the inside, we observed the causal structure: A space-like apparent horizon, a space-like singularity $r=0$ and a null-like Cauchy horizon. Even though we tune the gauge coupling parameter, the solution does not become a overcharged solution. Therefore, automatically the weak cosmic censorship is safe. On the other hand, the strong cosmic censorship seems not to be safe. The Cauchy horizon becomes singular due to the mass inflation, but the effect is much milder compared to the case of charged black hole. There may exist an observer inside the apparent horizon who is moving in the outgoing direction and that observer may not see the serious effects of mass inflation and can reach the Cauchy horizon. This is a quite different phenomena compared to the black hole case.

For the outside, we observed the dynamics of scalar perturbations. By adding an artificial scalar hair, the scalar field around the horizon is perturbed. For some over-charged examples, there remains a non-trivial scalar responses outside the horizon and there are observable imprints along the out-going direction. This can be interpreted as a non-trivial black hole solution with a charge cloud and this can have some meaning in the boundary conformal field theory.

To our knowledge, new observations in this paper are:
\begin{itemize}
\item We observed the \textit{dynamical} formation of black-branes.
\item We concisely investigated cosmic censorship of dynamical black-branes.
\item We see dynamics of scalar fields outside the black-brane with full back-reactions.
\end{itemize}
We may extend this analysis to other symmetries (topological black holes), other dimensions, or other theories (different couplings or potentials).

\appendix
\section*{Appendix: Convergence and constraint tests}
In this appendix, we demonstrate that our numerical code is converging (to a physical solution) when including the effects of
nontrivial scalar field, $V_0$ and charge. 

The results presented here are very similar to those presented in \cite{Hansen:2009kn} as could be expected, since the codes in those papers and this one are essentially identical. For this reason we refer to those papers and references therein, for further details of the inner workings of the code.

The initial conditions and computational domain for the tests in this Appendix are similar
to those used for making Figure~\ref{fig:causal}, i.e. the scalar pulse is between $10\le v\le 20$, $V_0 = -0.0001$ and charge $e = 1.0$. This scenario is non-trivial and thus a good test for the general convergence properties of our code. Our computational domain for this convergence tests is, as for Figure~\ref{fig:causal}, in the range $v=[0;100]$ and $u=[0;50]$.
This setup is one of the most complicated one that we have done in this
paper in that it leaves no trivial terms left in the evolution
equations and it is thus a good test for the convergence of our code.

Here we demonstrate the convergence of the code by comparing a series of simulations with varying numerical resolution. For the tests in this Appendix, we do a total of 6 simulations, with each simulation changing the base resolution.

To limit the number of plots, we concentrate on displaying convergence results along the line $u=30$, which (as can be seen by Figure~\ref{fig:causal}) is well inside the apparent horizon and quite close to the $r=0$ singularity. Thus, if we see convergence here, it is a fair indication that convergence requirements are satisfied throughout the whole of the computational domain. 

Along this line ($u=30$), we calculate the relative convergence between two simulations (one with a numerical resolution twice that of the other) relative to a simulation with very high resolution:
\begin{equation}
  \label{eq:xidef}
  \xi (x_N^i) \equiv \frac{|x_N^i - x_{2N}^i|}{ |x_{HighRes}^i |}
\end{equation}
where $x_N^i$ denotes the dynamic variable $x$ at the $i$-th grid
point of simulation with resolution $N$ and where $x_{HighRes}^i$
denotes the dynamic variable of the same $i$'th point for a
simulation with the highest numerical resolution done by us.
Obviously, this expression only makes sense for those $i$ points
that coincide in all simulations.

\begin{figure}
\begin{center}
\includegraphics[scale=0.5]{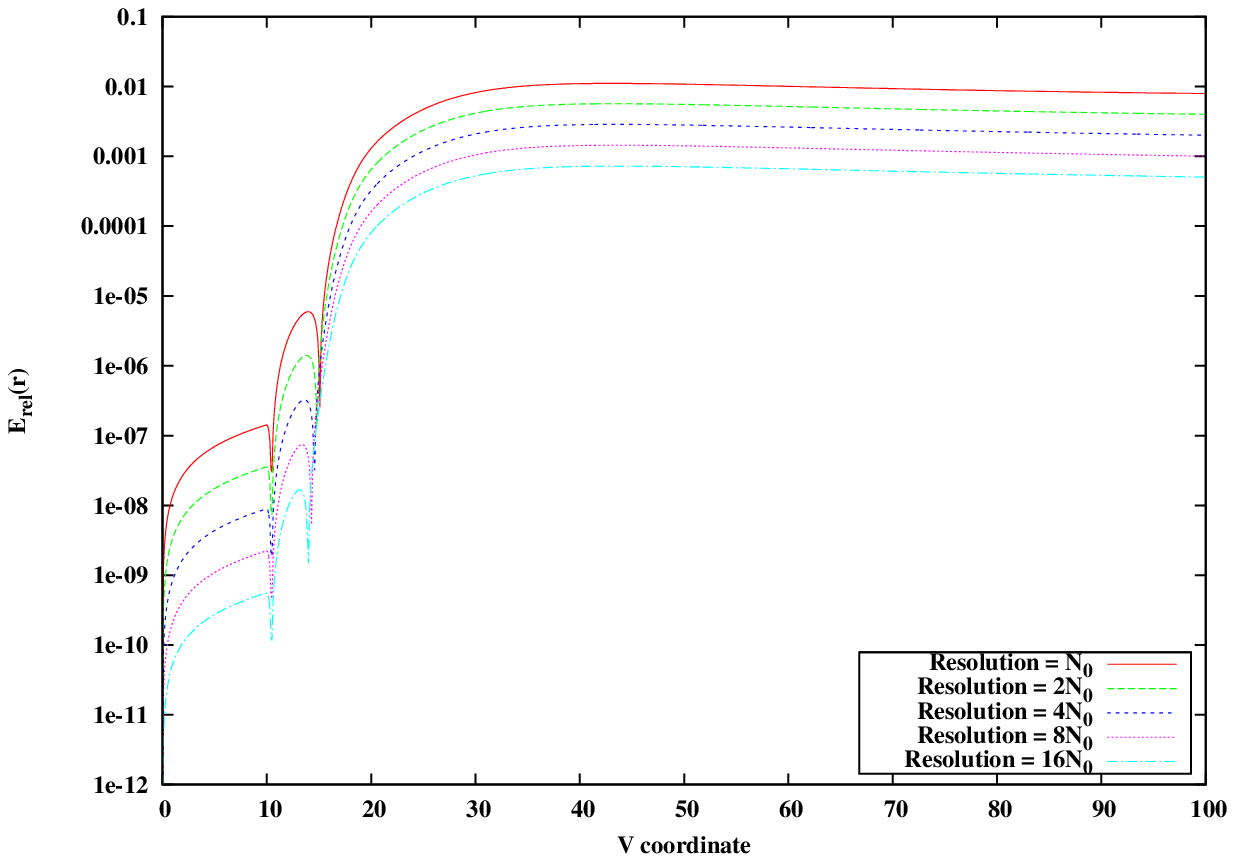}
\includegraphics[scale=0.5]{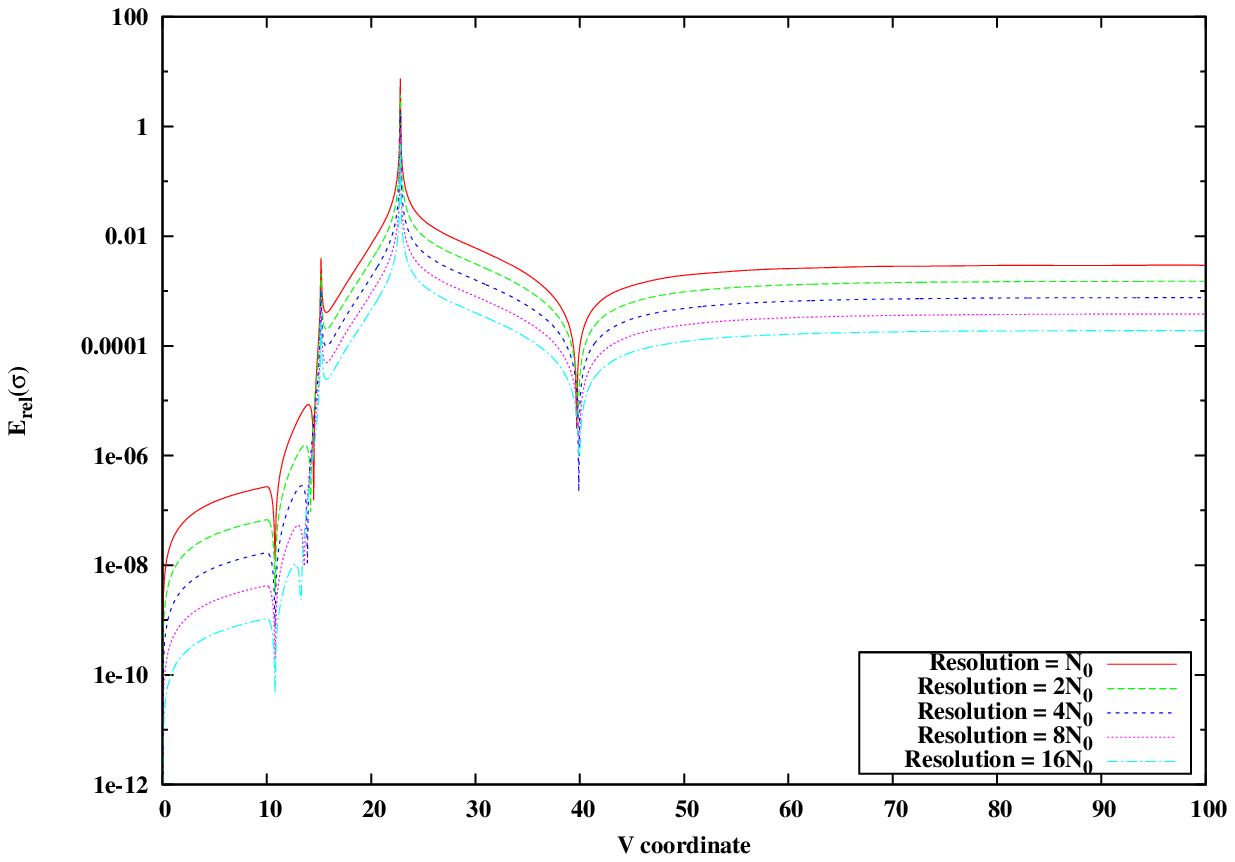}
\includegraphics[scale=0.5]{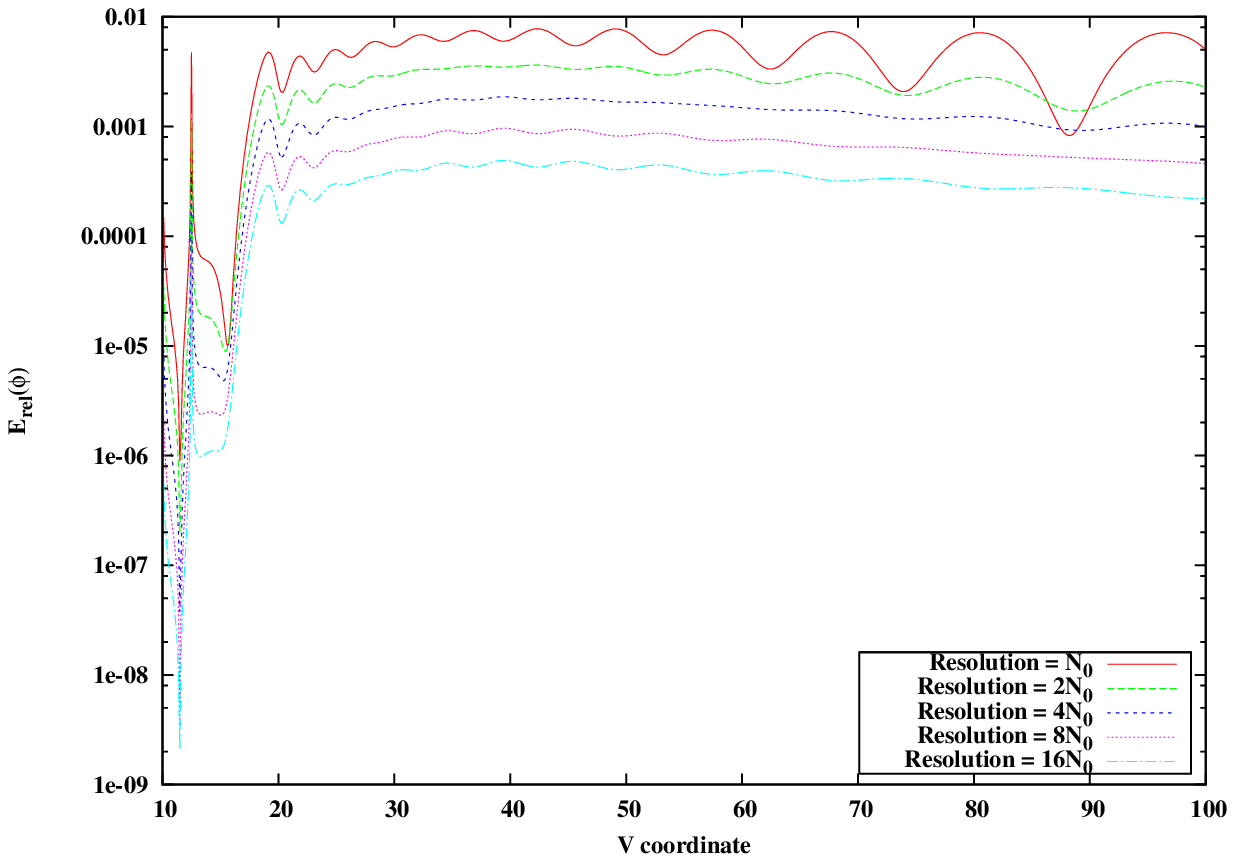}
\includegraphics[scale=0.5]{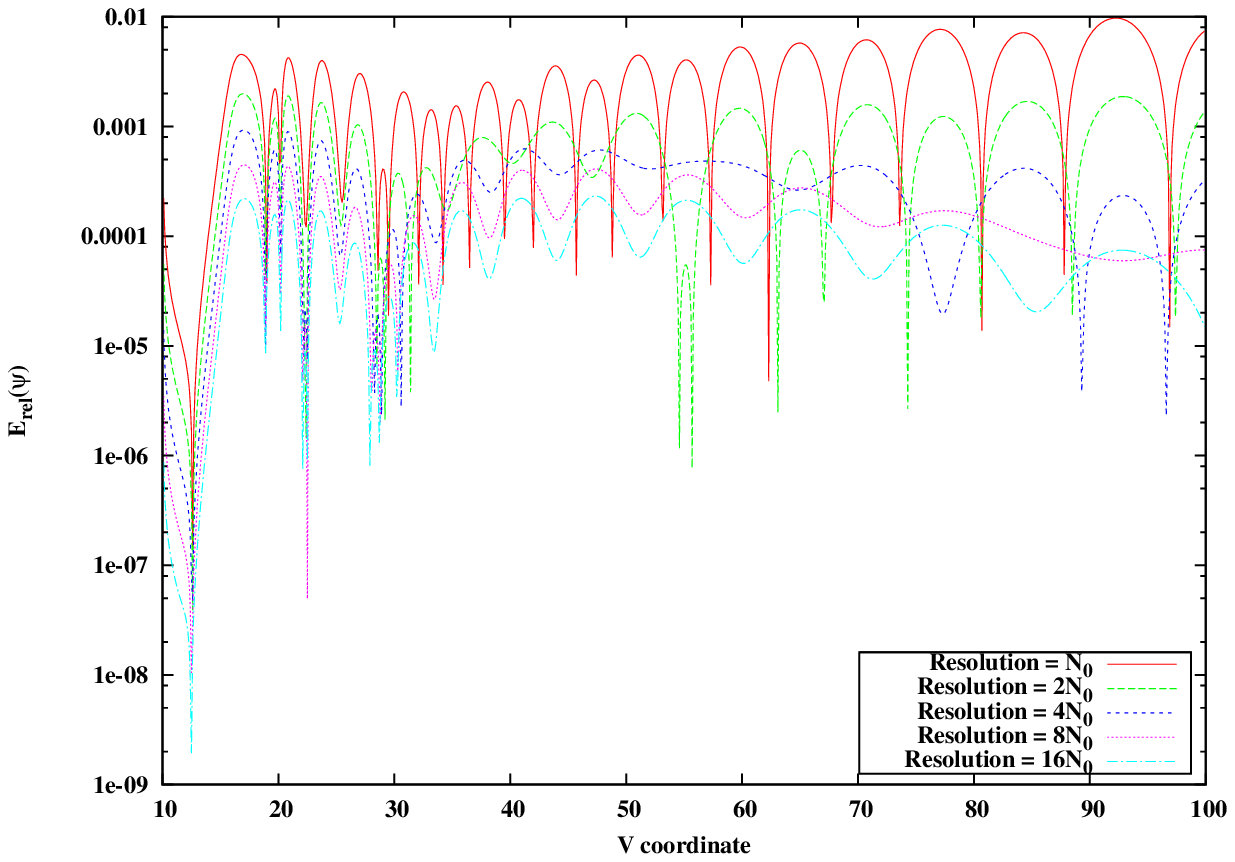}
\includegraphics[scale=0.5]{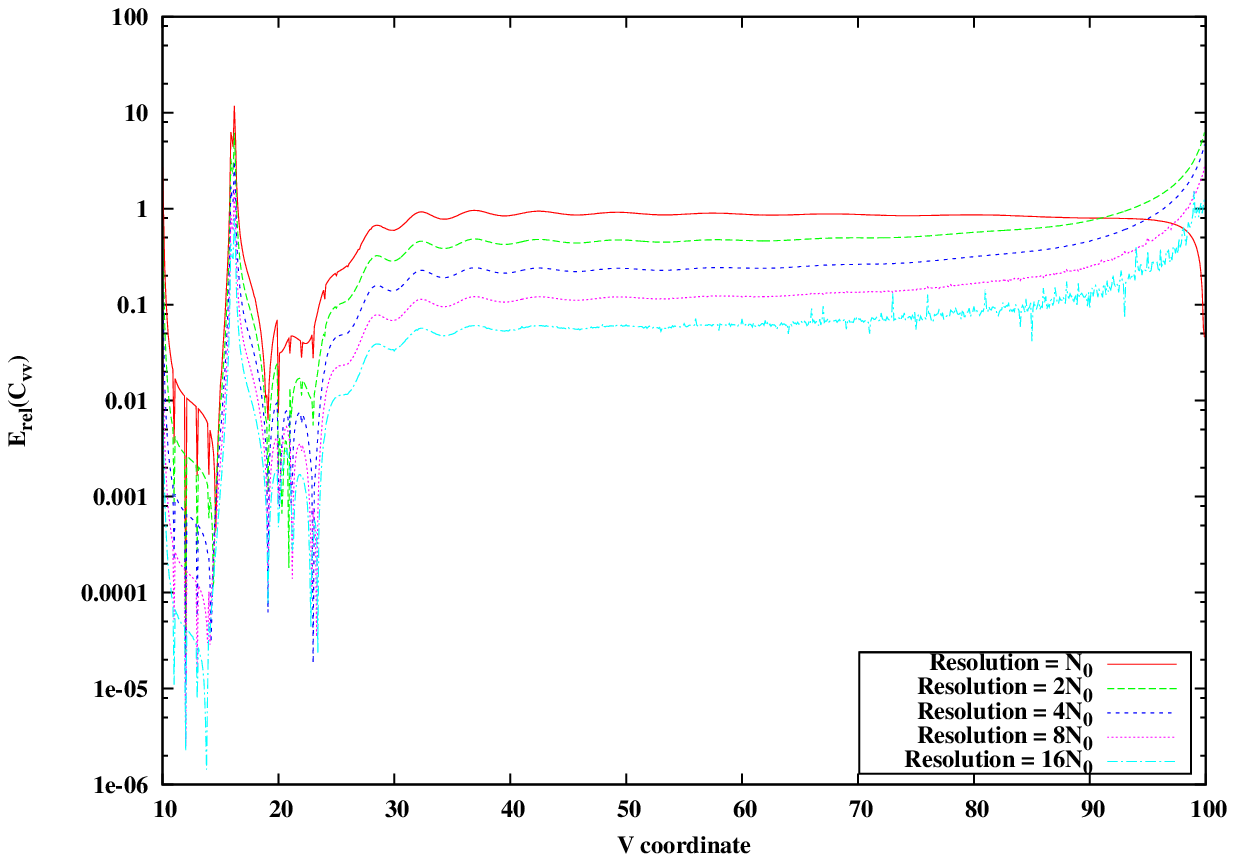}
\includegraphics[scale=0.5]{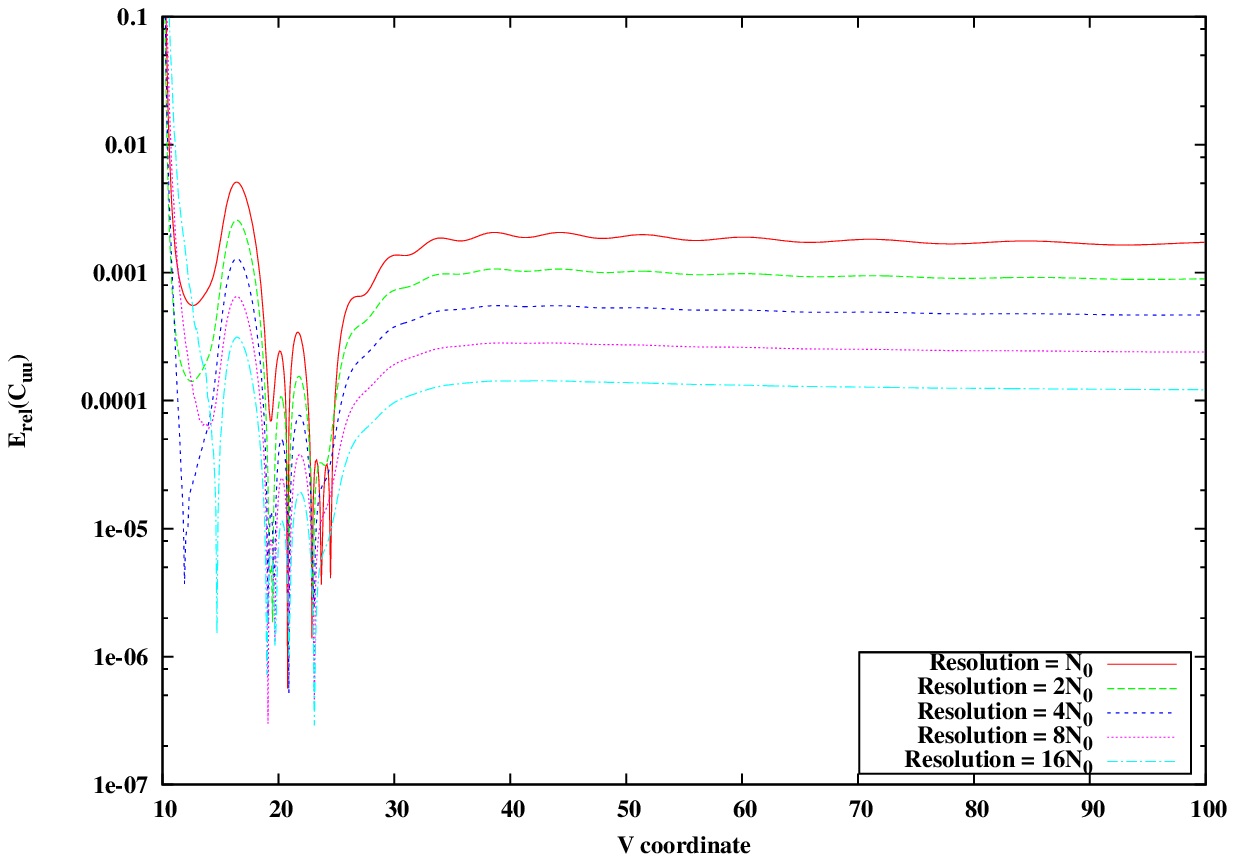}
\caption{\label{fig:convergence}Relative convergence of the dynamic variables and constraint
equations along line of $u=30$, see text for details.}
\end{center}
\end{figure}

The first four plots in Figure \ref{fig:convergence} show the relative convergence,
$\xi$, for the dynamic variables $r,\sigma,\Phi$ and $\Psi$
respectively. The lines in the figures are marked by their numerical resolution measured in terms of the most coarse resolution $N_0$, the high resolution simulation
used to calculate expression Equation~(\ref{eq:xidef}), has a numerical resolution of $32$ times the base resolution, i.e. $N=32N_0$.

From these figures, it is clearly seen that the four
dynamic variables are converging for simulations of increasing
resolution. Furthermore, since we plot the \textit{relative}
convergence of the dynamic variables, we see that the relative
change between the two highest resolution simulations show that
the variables change $0.1\%$ or less, which must be considered a
quite acceptable convergence. We note that a closer analysis of the data in
Figures \ref{fig:convergence} has revealed that the dynamic
variables are indeed converging with second order accuracy as was
expected based on analyses and tests performed in previous works
\cite{Hansen:2009kn}.

However, it is, of course, not enough to demonstrate that the
simulations are converging, they must also converge to a physical
solution, i.e. the residuals of the constraint equations must converge to zero. To
demonstrate this, we calculate the relative convergence of the
constraint equation residuals, (relative to the Einstein-tensor) in a similar way to Equation~(\ref{eq:xidef}):
\begin{equation}
  \label{eq:chidef}
  \chi (C_N^i) \equiv \frac{|C_N^i|}{ |G_{HighRes}^i |}
\end{equation}
where $C_N^i$ denotes the residual of the constraint equation,
($C_{uu}$ or $C_{vv}$), at the $i$'th point for simulation with
resolution $N$ and where $G_{HighRes}^i$ denotes the corresponding
Einstein-tensor component ($G_{uu}$ or $G_{vv}$ respectively) at
the same point.

The relative convergence of the residuals of the constraint
equations are demonstrated in the bottom two plots of Figure~\ref{fig:convergence}, where it is seen that they converge towards zero for
higher resolution simulations. This indicates that not only are
the numerical solutions converging for simulations of higher
resolution, but that they are indeed converging towards a physical
solution. 

Finally it should be noted that the convergence results presented
in this appendix are not the only convergence tests that we have
performed, they merely represent typical results of the
convergence behavior of the code. For all results presented in
this paper, we have performed a large number of simulations with
varying resolutions to ensure that the results had converged to
their physical solution.

\newpage

\section*{Acknowledgment}
DY, CP and BHL are supported by the National Research Foundation of Korea(NRF) funded by the Korea government(MEST, 2005-0049409) through the Center for Quantum Spacetime(CQUeST) of Sogang University. DY is supported by the JSPS Grant-in-Aid for Scientific Research (A) No.~21244033. CP was supported by Basic Science Research Program through the National Research Foundation of Korea(NRF) funded by the Ministry of Education, Science and Technology(2010-0022369). JH was supported in part by the research grant on {\it Computational Sciences and Research Hub} at KISTI and the APCTP Topical Research Program.


\begin{thebibliography}{200}

%\cite{Polchinski:1998rq}
\bibitem{Polchinski:1998rq}
  J.~Polchinski,
  ``{\it String theory. Vol. 1: An introduction to the bosonic string,}'' Cambridge University Press (1998); \\
  J.~Polchinski,
  ``{\it String theory. Vol. 2: Superstring theory and beyond,}'' Cambridge University Press (1998).

%\cite{Callan:1996dv}
\bibitem{Callan:1996dv}
  C.~G.~Callan and J.~M.~Maldacena,
%  ``D-brane approach to black hole quantum mechanics,''
  Nucl.\ Phys.\  B {\bf 472}, 591 (1996)
  [arXiv:hep-th/9602043].
  %%CITATION = NUPHA,B472,591;%%

%\cite{Maldacena:1997re}
\bibitem{Maldacena:1997re}
  J.~M.~Maldacena,
%  ``The large N limit of superconformal field theories and supergravity,''
  Adv.\ Theor.\ Math.\ Phys.\  {\bf 2}, 231 (1998)
  [Int.\ J.\ Theor.\ Phys.\  {\bf 38}, 1113 (1999)]
  [arXiv:hep-th/9711200].

%\cite{Gubser:1998bc}
\bibitem{Gubser:1998bc} 
  S.~S.~Gubser, I.~R.~Klebanov and A.~M.~Polyakov,
  %``Gauge theory correlators from noncritical string theory,''
  Phys.\ Lett.\ B {\bf 428}, 105 (1998)
  [hep-th/9802109].
  %%CITATION = HEP-TH/9802109;%%
  %5268 citations counted in INSPIRE as of 21 May 2013
  
  %\cite{Witten:1998qj}
\bibitem{Witten:1998qj} 
  E.~Witten,
  %``Anti-de Sitter space and holography,''
  Adv.\ Theor.\ Math.\ Phys.\  {\bf 2}, 253 (1998)
  [hep-th/9802150].
  %%CITATION = HEP-TH/9802150;%%
  %6007 citations counted in INSPIRE as of 21 May 2013
  
%\cite{Hamade:1995ce}
\bibitem{Hamade:1995ce}
  R.~S.~Hamade and J.~M.~Stewart,
%  ``The spherically symmetric collapse of a massless scalar field,''
  Class.\ Quant.\ Grav.\  {\bf 13}, 497 (1996)
  [arXiv:gr-qc/9506044].
  %%CITATION = CQGRD,13,497;%%
  
%\cite{doublenull}
\bibitem{doublenull}
  T.~Piran and A.~Strominger, Phys.\ Rev.\ D {\bf 48}, 4729 (1993) [arXiv:hep-th/9304148];\\
  R.~Parentani and T.~Piran, Phys.\ Rev.\ Lett {\bf 73}, 2805 (1994) [arXiv:hep-th/9405007];\\
  T.~Chiba and J.~Soda, Prog.\ Theor.\ Phys.\  {\bf 96}, 567 (1996) [arXiv:gr-qc/9603056]; \\
  S.~Ayal and T.~Piran, Phys.\ Rev.\  D {\bf 56}, 4768 (1997) [arXiv:gr-qc/9704027];\\
  S.~Hod and T.~Piran, Phys.\ Rev.\ Lett {\bf 81}, 1554 (1998) [arXiv:gr-qc/9803004];\\
  S.~Hod and T.~Piran, Gen.\ Rel.\ Grav.\  {\bf 30}, 1555 (1998) [arXiv:gr-qc/9902008];\\
  E.~Sorkin and T.~Piran, Phys.\ Rev.\ D {\bf 63}, 084006 (2001) [arXiv:gr-qc/0009095];\\
  E.~Sorkin and T.~Piran, Phys.\ Rev.\  D {\bf 63}, 124024 (2001) [arXiv:gr-qc/0103090];\\
  Y.~Oren and T.~Piran, Phys.\ Rev.\ D {\bf 68}, 044013 (2003) [arXiv:gr-qc/0306078];\\
  P.~P.~Avelino, A.~J.~S.~Hamilton and C.~A.~R.~Herdeiro, Phys.\ Rev.\  D {\bf 79}, 124045 (2009) [arXiv:0904.2669 [gr-qc]]; \\
  A.~Borkowska, M.~Rogatko and R.~Moderski, Phys.\ Rev.\  D {\bf 83}, 084007 (2011) [arXiv:1103.4808 [hep-th]];\\
  A.~Nakonieczna, M.~Rogatko and R.~Moderski,  Phys.\ Rev.\ D {\bf 86}, 044043 (2012)  [arXiv:1209.1203 [hep-th]];\\
  A.~Nakonieczna and M.~Rogatko,  arXiv:1209.3614 [hep-th].

%\cite{Hansen:2009kn}
\bibitem{Hansen:2009kn}
  J.~Hansen, A.~Khokhlov and I.~Novikov, Phys.\ Rev.\  D {\bf 71}, 064013 (2005) [arXiv:gr-qc/0501015];\\
  A.~Doroshkevich, J.~Hansen, I.~Novikov and A.~Shatskiy, arXiv:0812.0702 [gr-qc];\\
  J.~Hansen, D.~Hwang and D.~Yeom,
  %``Dynamics of false vacuum bubbles: Beyond the thin shell approximation,''
  JHEP {\bf 0911}, 016 (2009)
  [arXiv:0908.0283 [gr-qc]];\\
  %%CITATION = ARXIV:0908.0283;%%
  A.~Doroshkevich, J.~Hansen, D.~Novikov, I.~Novikov and A.~Shatskiy, Phys.\ Rev.\  D {\bf 81}, 124011 (2010)
  [arXiv:0908.1300 [gr-qc]].

%\cite{Hong:2008mw}
\bibitem{Hong:2008mw}
  S.~E.~Hong, D.~Hwang, E.~D.~Stewart and D.~Yeom,
  %``The causal structure of dynamical charged black holes,''
  Class.\ Quant.\ Grav.\  {\bf 27}, 045014 (2010)
  [arXiv:0808.1709 [gr-qc]].
  %%CITATION = CQGRD,27,045014;%%

%\cite{Hwang:2011mn}
\bibitem{Hwang:2011mn}
  D.~Hwang, H.~-B.~Kim and D.~Yeom,
  %``Dynamical formation and evolution of (2+1)-dimensional charged black holes,''
  Class.\ Quant.\ Grav.\  {\bf 29}, 055003 (2012)
  [arXiv:1105.1371 [gr-qc]].
  %%CITATION = ARXIV:1105.1371;%%


%\cite{Hwang:2010aj}
\bibitem{Hwang:2010aj}
  D.~Yeom,
  %``Generation of a bubble universe and the information loss problem,''
  arXiv:0912.0068 [gr-qc];\\
  %%CITATION = ARXIV:0912.0068;%%
  D.~Hwang and D.~Yeom,
  %``Generation of a bubble universe using a negative energy bath,''
  Class.\ Quant.\ Grav.\  {\bf 28}, 155003 (2011)
  [arXiv:1010.3834 [gr-qc]];\\
  %%CITATION = ARXIV:1010.3834;%%
  D.~Hwang and D.~Yeom,
  %``Responses of the Brans-Dicke field due to gravitational collapses,''
  Class.\ Quant.\ Grav.\  {\bf 27}, 205002 (2010)
  [arXiv:1002.4246 [gr-qc]];\\
  %%CITATION = ARXIV:1002.4246;%%
  D.~Hwang and D.~Yeom,
  %``Internal structure of charged black holes,''
  Phys.\ Rev.\ D {\bf 84}, 064020 (2011)
  [arXiv:1010.2585 [gr-qc]];\\
  %%CITATION = ARXIV:1010.2585;%%
  D.~Hwang, B.~-H.~Lee and D.~Yeom,
  %``Mass inflation in f(R) gravity: A Conjecture on the resolution of the mass inflation singularity,''
  JCAP {\bf 1112}, 006 (2011)
  [arXiv:1110.0928 [gr-qc]];\\
  %%CITATION = ARXIV:1110.0928;%%
  B.~-H.~Lee and D.~Yeom,
  %``Nucleation and Evolution of False Vacuum Bubbles in Scalar-Tensor Gravity,''
  Nuovo Cim.\ C {\bf 36}, S1, 79 (2013)
  [arXiv:1111.0139 [gr-qc]];\\
  D.~Hwang, B.~-H.~Lee and D.~Yeom,
  %``Is the firewall consistent?: Gedanken experiments on black hole complementarity and firewall proposal,''
  JCAP {\bf 1301}, 005 (2013)
  [arXiv:1210.6733 [gr-qc]];\\
  %%CITATION = ARXIV:1210.6733;%%
  D.~Hwang, F.~G.~Pedro and D.~Yeom,
  arXiv:1306.6687 [hep-th].
  %%CITATION = ARXIV:1306.6687;%%


%\cite{Hwang:2012pj}
\bibitem{Hwang:2012pj} 
  D.~Hwang, B.~-H.~Lee, W.~Lee and D.~Yeom,
  %``Bubble collision with gravitation,''
  JCAP {\bf 1207}, 003 (2012)
  [arXiv:1201.6109 [gr-qc]].
  %%CITATION = ARXIV:1201.6109;%%

%\cite{Hartnoll:2008vx}
\bibitem{Hartnoll:2008vx} 
  S.~A.~Hartnoll, C.~P.~Herzog and G.~T.~Horowitz,
  %``Building a Holographic Superconductor,''
  Phys.\ Rev.\ Lett.\  {\bf 101}, 031601 (2008)
  [arXiv:0803.3295 [hep-th]];\\
  %%CITATION = ARXIV:0803.3295;%%
  %519 citations counted in INSPIRE as of 21 May 2013
%\cite{Hartnoll:2008kx}
%\bibitem{Hartnoll:2008kx} 
  S.~A.~Hartnoll, C.~P.~Herzog and G.~T.~Horowitz,
  %``Holographic Superconductors,''
  JHEP {\bf 0812}, 015 (2008)
  [arXiv:0810.1563 [hep-th]].
  %%CITATION = ARXIV:0810.1563;%%
  %377 citations counted in INSPIRE as of 21 May 2013
  
  
  
%\cite{Erlich:2005qh}
\bibitem{Erlich:2005qh} 
  J.~Erlich, E.~Katz, D.~T.~Son and M.~A.~Stephanov,
  %``QCD and a holographic model of hadrons,''
  Phys.\ Rev.\ Lett.\  {\bf 95}, 261602 (2005)
  [hep-ph/0501128]; \\
  %%CITATION = HEP-PH/0501128;%%
  %546 citations counted in INSPIRE as of 21 May 2013
  %\cite{Karch:2006pv}
%\bibitem{Karch:2006pv} 
  A.~Karch, E.~Katz, D.~T.~Son and M.~A.~Stephanov,
  %``Linear confinement and AdS/QCD,''
  Phys.\ Rev.\ D {\bf 74}, 015005 (2006)
  [hep-ph/0602229]; \\
  %%CITATION = HEP-PH/0602229;%%
  %453 citations counted in INSPIRE as of 21 May 2013
%\cite{Park:2011ab}
%\bibitem{Park:2011ab} 
  C.~Park, B.~-H.~Lee and S.~Shin,
  %``Holographic Meson Spectra in the Dense Medium with Chiral Condensate,''
  Phys.\ Rev.\ D {\bf 85}, 106005 (2012)
  [arXiv:1112.2177 [hep-th]].
  %%CITATION = ARXIV:1112.2177;%%

%\cite{Lee:2009bya}
\bibitem{Lee:2009bya} 
  B.~-H.~Lee, C.~Park and S.~-J.~Sin,
  %``A Dual Geometry of the Hadron in Dense Matter,''
  JHEP {\bf 0907}, 087 (2009)
  [arXiv:0905.2800 [hep-th]]; \\
  %%CITATION = ARXIV:0905.2800;%%
  %17 citations counted in INSPIRE as of 21 May 2013
%\cite{Park:2009nb}
%\bibitem{Park:2009nb} 
  C.~Park,
  %``The Dissociation of a heavy meson in the quark medium,''
  Phys.\ Rev.\ D {\bf 81}, 045009 (2010)
  [arXiv:0907.0064 [hep-ph]]; \\
  %%CITATION = ARXIV:0907.0064;%%
  %10 citations counted in INSPIRE as of 21 May 2013
%\cite{Park:2011zp}
%\bibitem{Park:2011zp} 
  C.~Park,
  %``Holographic Symmetry Energy of the Nuclear Matter,''
  Phys.\ Lett.\ B {\bf 708}, 324 (2012)
  [arXiv:1112.0386 [hep-th]].
  %%CITATION = ARXIV:1112.0386;%%
  %3 citations counted in INSPIRE as of 21 May 2013

  
%\cite{Kachru:2008yh}
\bibitem{Kachru:2008yh} 
  S.~Kachru, X.~Liu and M.~Mulligan,
  %``Gravity Duals of Lifshitz-like Fixed Points,''
  Phys.\ Rev.\ D {\bf 78}, 106005 (2008)
  [arXiv:0808.1725 [hep-th]];\\
  %%CITATION = ARXIV:0808.1725;%%
  %347 citations counted in INSPIRE as of 21 May 2013
  %\cite{Taylor:2008tg}
%\bibitem{Taylor:2008tg} 
  M.~Taylor,
  %``Non-relativistic holography,''
  arXiv:0812.0530 [hep-th];\\
  %%CITATION = ARXIV:0812.0530;%%
  %180 citations counted in INSPIRE as of 21 May 2013
 %\cite{Goldstein:2009cv}
%\bibitem{Goldstein:2009cv} 
  K.~Goldstein, S.~Kachru, S.~Prakash and S.~P.~Trivedi,
  %``Holography of Charged Dilaton Black Holes,''
  JHEP {\bf 1008}, 078 (2010)
  [arXiv:0911.3586 [hep-th]];\\
  %%CITATION = ARXIV:0911.3586;%%
  %151 citations counted in INSPIRE as of 21 May 2013 
  %\cite{Charmousis:2010zz}
%\bibitem{Charmousis:2010zz} 
  C.~Charmousis, B.~Gouteraux, B.~S.~Kim, E.~Kiritsis and R.~Meyer,
  %``Effective Holographic Theories for low-temperature condensed matter systems,''
  JHEP {\bf 1011}, 151 (2010)
  [arXiv:1005.4690 [hep-th]];\\
  %%CITATION = ARXIV:1005.4690;%%
  %138 citations counted in INSPIRE as of 21 May 2013
  %\cite{Goldstein:2010aw}
%\bibitem{Goldstein:2010aw} 
  K.~Goldstein, N.~Iizuka, S.~Kachru, S.~Prakash, S.~P.~Trivedi and A.~Westphal,
  %``Holography of Dyonic Dilaton Black Branes,''
  JHEP {\bf 1010}, 027 (2010)
  [arXiv:1007.2490 [hep-th]];\\
  %%CITATION = ARXIV:1007.2490;%%
  %73 citations counted in INSPIRE as of 21 May 2013
  %\cite{Kulkarni:2012re}
%\bibitem{Kulkarni:2012re} 
  S.~Kulkarni, B.~-H.~Lee, C.~Park and R.~Roychowdhury,
  %``Non-conformal Hydrodynamics in Einstein-dilaton Theory,''
  JHEP {\bf 1209}, 004 (2012)
  [arXiv:1205.3883 [hep-th]];\\
  %%CITATION = ARXIV:1205.3883;%%
  %2 citations counted in INSPIRE as of 21 May 2013
%\cite{Park:2012cu}
%\bibitem{Park:2012cu} 
  C.~Park,
  %``Membrane paradigm in the Einstein-dilaton theory,''
  arXiv:1209.0842 [hep-th];\\
  %%CITATION = ARXIV:1209.0842;%%
  %4 citations counted in INSPIRE as of 21 May 2013
%\cite{Kulkarni:2012in}
%\bibitem{Kulkarni:2012in} 
  S.~Kulkarni, B.~-H.~Lee, J.~-H.~Oh, C.~Park and R.~Roychowdhury,
  %``Transports in non-conformal holographic fluids,''
  JHEP {\bf 1303}, 149 (2013)
  [arXiv:1211.5972 [hep-th]].
  %%CITATION = ARXIV:1211.5972;%%
  %1 citations counted in INSPIRE as of 21 May 2013


%\cite{wald}
\bibitem{wald}
  R.~M.~Wald, ``{\it General relativity,}'' University of Chicago Press (1984).

%\cite{Hawking:1973uf}
\bibitem{Hawking:1973uf}
  S.~W.~Hawking and G.~F.~R.~Ellis,
  ``{\it The large scale structure of space-time,}'' Cambridge University Press (1973).



%\cite{Poisson:1989zz}
\bibitem{Poisson:1989zz}
  E.~Poisson and W.~Israel,
  %``Inner-horizon instability and mass inflation in black holes,''
  Phys.\ Rev.\ Lett.\  {\bf 63}, 1663 (1989); \\
  %%CITATION = PRLTA,63,1663;%%
  E.~Poisson and W.~Israel,
  %``Internal structure of black holes,''
  Phys.\ Rev.\  D {\bf 41}, 1796 (1990).
  %%CITATION = PHRVA,D41,1796;%%

%\cite{Poisson:1997my}
\bibitem{Poisson:1997my}
  E.~Poisson, [arXiv:gr-qc/9709022].

%\cite{Price}
\bibitem{Price}
R.H. Price, Phys.\ Rev.\  D {\bf 5}, 2419 (1972);\\
R.H. Price, Phys.\ Rev.\  D {\bf 5}, 2439 (1972).

%\cite{Ori}
\bibitem{Ori}
  A.~Ori,  Phys.\ Rev.\ Lett.\  {\bf 67}, 789 (1991); \\
  A.~Ori,  Phys.\ Rev.\ Lett.\  {\bf 68}, 2117 (1992).



\end{thebibliography}
\end{document}